%% file: main.tex
\documentclass[sigconf]{acmart}

\def\BibTeX{{\rm B\kern-.05em{\sc i\kern-.025em b}\kern-.08emT\kern-.1667em\lower.7ex\hbox{E}\kern-.125emX}}
    
\copyrightyear{2019}
\acmYear{2019}
\setcopyright{acmlicensed}
\acmConference[The Web Conference '19]{The Web Conference '19: Data Science for Social Good Workshop}{May 14, 2019}{San Francisco, CA}
\acmBooktitle{The Web Conference '19: Data Science for Social Good Workshop, May 14, 2019, San Francisco, CA}
\acmPrice{15.00}
\acmDOI{10.1145/1122445.1122456}
\acmISBN{978-1-4503-9999-9/18/06}

\usepackage{times}  
\usepackage{helvet}  
\usepackage{courier}  
\usepackage{url}  
\usepackage{graphicx}  
\usepackage{color-edits}
\usepackage{amsmath}
\usepackage{balance}
\usepackage{subcaption}
\usepackage{mwe}
\usepackage{booktabs}
\usepackage{multirow}

\addauthor{ab}{blue}  
\addauthor{ad}{red}
\addauthor{ef}{green}
\addauthor{ll}{cyan}

\begin{document}

\title[Perils and Challenges of Social Media and Election Manipulation]{Perils and Challenges of Social Media and Election Manipulation Analysis: The 2018 US Midterms}
\author[A. Deb]{Ashok Deb}
\affiliation{%
  \institution{USC Information Sciences Institute}
  \city{Marina del Rey}
  \state{CA}
  \postcode{90292}
}
\email{ashok@isi.edu}

\author[L. Luceri]{Luca Luceri}
\additionalaffiliation{%
  \institution{USC Information Sciences Institute}
  \city{Marina del Rey}
  \state{CA}
}
\affiliation{%
  \institution{University of Applied Sciences and Arts of Southern Switzerland, and University of Bern}
  \city{Manno}
  \state{Switzerland}
  \postcode{6928}
}
\email{luca.luceri@supsi.ch}

\author[A. Badaway]{Adam Badaway}
\affiliation{%
  \institution{USC Information Sciences Institute}
  \city{Marina del Rey}
  \state{CA}
  \postcode{90292}
}
\email{badawy@isi.edu}

\author[E. Ferrara]{Emilio Ferrara}
\affiliation{%
  \institution{USC Information Sciences Institute}
  \city{Marina del Rey}
  \state{CA}
  \postcode{90292}
}
\email{emiliofe@usc.edu}



\begin{abstract}

One of the hallmarks of a free and fair society is the ability to conduct a peaceful and seamless transfer of power from one leader to another. Democratically, this is measured in a citizen population's trust in the electoral system of choosing a representative government. In view of the well documented issues of the 2016 US Presidential election, we conducted an in-depth analysis of the 2018 US Midterm elections looking specifically for voter fraud or suppression. The Midterm election occurs in the middle of a 4 year presidential term. For the 2018 midterms, 35 Senators and all the 435 seats in the House of Representatives were up for re-election, thus, every congressional district and practically every state had a federal election. In order to collect election related tweets, we analyzed Twitter during the month prior to, and the two weeks following, the November 6, 2018 election day. In a targeted analysis to detect statistical anomalies or election interference, we identified several biases that can lead to wrong conclusions. Specifically, we looked for divergence between actual voting outcomes and instances of the \#ivoted hashtag on the election day. This analysis highlighted three states of concern: New York, California, and Texas. We repeated our analysis discarding malicious accounts, such as social bots. Upon further inspection and against a backdrop of collected general election-related tweets, we identified some confounding factors, such as population bias, or bot and political ideology inference, that can lead to false conclusions. We conclude by providing an in-depth discussion of the perils and challenges of using social media data to explore questions about election manipulation.

\end{abstract}

\maketitle

\input{src/introduction.tex}

\input{src/background.tex}

\input{src/related_work.tex}
\input{src/data.tex}

\input{src/methodology.tex}

\input{src/results.tex}
\input{src/discussion.tex}

\input{src/conclusion.tex}

\balance
\newpage 
\footnotesize{
\textbf{Acknowledgements}. 
The authors gratefully acknowledge support by the Air Force Office of Scientific Research (award \#FA9550-17-1-0327). 
L. Luceri is funded by the Swiss National Science Foundation (SNSF) via the CHIST-ERA project \textit{UPRISE-IoT}.
}
\bibliographystyle{ACM-Reference-Format}
\bibliography{ref}

\end{document}

%% file: src/introduction.tex
\section{Introduction}

Inherent bias of drawing conclusions from political polls stretch back to the famous headline of "Dewey Defeats Truman" in the 1948 US Presidential election \cite{mosteller1949pre}. Confounding factors that led to false conclusions in the 1948 election included telephone surveys which did not use robust statistical methods and an under-sampling of Truman supporters. Likewise, in 2016, many political pundits underestimated the likelihood that Donald Trump would be elected as President of the United States. 
The research community demonstrated a strong interest in studying  social media to get a better understanding of how the 2016 events unfolded.  
Numerous studies  concluded that social media can be a vehicle for political manipulation, citing factors such as the effect of fake news and disinformation~\cite{persily20172016, howard2017junk, shu2017fake, vosoughi2018spread, badawy2018falls, guess2019less, bovet2019influence, scheufele2019science, Grinberg2019}, bots~\cite{bessi2016social, woolley2017computational, varol2017online, monsted2017evidence, boichak2018automated, shao2018spread, yang2019arming}, polarization \cite{bail2018exposure, azzimonti2018social}, etc.

Research also suggests that social media data comes with  significant biases that limit the ability to forecast offline events, e.g., the outcomes of political elections \cite{metaxas2011not, gayo2011don, gayo2011limits, gayo2012wanted, gayo2012no, gayo2013meta}, or public health issues \cite{lazer2014parable, allem2017cigarette,  williams2017crime}.
Despite these well documented issues and challenges, social media are frequently relied upon and referred to as a trusted source of information to speculate about, or try to explain, offline events. One such example is the recent 2018 US Midterm elections where widespread claims of voter fraud and voter suppression appeared in the news, often based on social media reports and accounts.

In this paper, we seek to understand whether it is possible to use Twitter as a sensor to estimate the expected amount of votes generated by each state.
We propose an undertaking in which we use the tweets with the hashtag \#ivoted on the election day as a proxy for actual votes. At first, this seemed like a promising research direction, as tweet volumes and vote counts correlated well for 47 of the 50 states in America. We also considered if this would be a useful approach to detecting voting issues like fraud or suppression, for example by isolating statistical anomalies in  estimated and observed volumes. To get a sense of expected tweet volume, we carried out the same analysis against general keywords related to the midterm election from a month before election day through two weeks after the election. We also considered how bots may have had an influence on election manipulation narratives by measuring their activity in the social media discourse. 
We finally applied a political ideology inference technique and tested it to see how well it compared to an external source of polls data.

The conclusions from our analysis are complex, and this work is meant as a note of caution about the risks of using social media analysis to infer political election manipulation such as voter fraud and voter suppression.

\subsection{Contributions of this work}
After exploring multiple Twitter data sets and two external sources (vote counts and Gallup), we came to the following contributions:
\begin{itemize}
    \item We explored how social media analysis carries a lot of risks involved mainly with population bias, data collection bias, lack of location-specific data, separation of bots (and organizations) from humans, information verification and fact-checking, and lastly assigning political ideology. 
    \item We saw a significant difference in the removal of retweets in our analysis as compared with including them. However, the effect was isolated to one particular state, Texas, indicating that the sensitivity of this effect could be a factor of location.
    \item There is a significant difference between people's reported political ideologies using a source like Gallup versus that can be inferred on social media. It is not possible to know if this is due to limitations of political inference algorithms, confounders, population representation biases, or else. 
    \item In the two states (NY \& TX) where there was a statistically significant discrepancy between vote counts and instances of self-reported voting via \#ivoted hashtags, we found only limited anecdotal evidence of tweets reporting issues of voter fraud or suppression. The divergence can possibly be explained by confounding factors, locality and selection bias, or social influence of particular candidates in those states (e.g.,  Alexandria Ocasio-Cortez in NY and Beto O'Rourke in TX).
\end{itemize}

%% file: src/background.tex
\section{Background}

The US Midterm elections were held on 6 November, 2018. They are referred to as mid-term elections because they occur in the middle of a presidential term. Senators serve for 6 years, thus, every 2 years, nearly a third of the Senators are up for re-election. 
The Senate is divided into 3 classes, depending on which year they were elected. Class I was elected in 2012 and are up for re-election in 2018. 

For 2018, 35 Senators out of a total of 100 senators in the 115th Congress will be up for re-election. Of the 35 senators up for election, 33 are in Senate Class I and two are Senators who vacated, whereas 15 are in what is to be considered contentious races. The 33 Class I are 30 (23 Democrats (D), 5 Republicans (R), 2 Independents (I)) up for re-election and 3 Republicans (R) who are retiring.
Details on the Senate seats up for re-election are in Table \ref{tab:senators}. 
Additionally, all 535 House of Representative seats are up for re-election every 2 years. Excluded from our analysis are the non-voting delegates for DC and the US Territories.

\begin{table}[t]
\centering \small
\caption{US Senate Seats Up for Election in 2018
\label{tab:senators}} 
\begin{tabular}{|l|c|c|l|}
        	\hline
           
        		Incumbent& State& Party& Status\\ 
        	\hline		
			
Tammy Baldwin	&	WI	&	D	&	Contested	\\
John Barraso	&	WY	&	R	&	Safe	\\
Sherrod Brown	&	OH	&	D	&	Contested	\\
Maria Cantrell	&	WA	&	D	&	Safe	\\
Ben Cardin	&	MD	&	D	&	Safe	\\
Tom Carper	&	DE	&	D	&	Safe	\\
Bob Casey	&	PA	&	D	&	Safe	\\
Bob Corker	&	TN	&	R	&	Retiring	\\
Ted Cruz	&	TX	&	R	&	Contested	\\
Joe Donnelly	&	IN	&	D	&	Contested	\\
Dianne Feinstein	&	CA	&	D	&	Safe	\\
Deb Fischer	&	NE	&	R	&	Safe	\\
Jeff Flake	&	AZ	&	R	&	Retiring	\\
Kirsten Gillibrand	&	NY	&	D	&	Safe	\\
Orrin Hatch	&	UT	&	R	&	Retiring	\\
Martin Heinrich	&	NM	&	D	&	Safe	\\
Heidi Heitkamp	&	ND	&	D	&	Contested	\\
Dean Heller	&	NV	&	R	&	Contested	\\
Mazie Hirono	&	HI	&	D	&	Safe	\\
Cindy Hyde-Smith	&	MS	&	R	&	Contested	\\
Tim Kaine	&	VA	&	D	&	Safe	\\
Angus King	&	ME	&	I	&	Safe	\\
Amy Klobuchar	&	MN	&	D	&	Safe	\\
Joe Manchin	&	WV	&	D	&	Contested	\\
Claire McCaskill	&	MO	&	D	&	Contested	\\
Bob Menendez	&	NJ	&	D	&	Contested	\\
Chris Murphy	&	CT	&	D	&	Safe	\\
Bill Nelson	&	FL	&	D	&	Contested	\\
Bernie Sanders	&	VT	&	I	&	Safe	\\
Tina Smith	&	MN	&	D	&	Contested	\\
Debbie Stabenow	&	MI	&	D	&	Safe	\\
Jon Tester	&	MT	&	D	&	Contested	\\
Elizabeth Warren	&	MA	&	D	&	Safe	\\
Sheldon Whitehouse	&	RI	&	D	&	Safe	\\
Roger Wicker	&	MS	&	R	&	Safe	\\
             \hline
\end{tabular}
\end{table}

%% file: src/related_work.tex
\section{Related Work}

Since the 2016 US Presidential election, there has been a big spotlight on the sovereignty of the US election system. The \textit{Bot Disclosure and Accountability Act} of 2018\footnote{\url{https://www.congress.gov/bill/115th-congress/senate-bill/3127/text}} gave clear guidelines for what has to be disclosed by social media companies. The article \textit{The Rise of Social Bots} \cite{ferrara2016rise} brought awareness to the issue of social bots in social media platforms. In \cite{bessi2016social}, Bessi \& Ferrara focused on social bots detection within the online discussion related to the 2016 presidential election. Other than characterizing the behavioral differences between humans and bots, there was not an in-depth analysis of any malicious intent. In this paper, we address the potential malicious activity in online political discussion along the lines of voter fraud, voter suppression, political misinformation, and then report on the biases we found.

\subsection{Voting Issues}
Concerns related to voter fraud took center stage after the 2000 US Presidential election, where it was argued that the candidate with the most votes lost and the Supreme Court decided the winner \cite{minnite2017myth}. 
Since then, a host of public debate, congressional testimony, and several new laws passed, such as the Help America Vote Act \cite{kim2003help}, which surprisingly needed to happened after the National Voter Registration Act of 1993 (NVRA).\footnote{\url{https://www.justice.gov/crt/about-national-voter-registration-act}} The effects of the NVRA were researched by \citet{highton1998estimating}, who concluded that provisions in the NVRA would increase voter turnout by 4.7\%-8.7\% and that purging voter rolls of those who had not voted in the last two years would have a 2\% effect. Lastly, they identified the two most vulnerable non-voting groups to be those under the age of 30 and those who moved within 2 years of an election \cite{highton1998estimating}. 

Moreover, it has been argued that the current US voter registration has a minimal impact on registration and that there is marginal value in any updated laws \cite{highton2004voter}. 
Therefore, the main concern argued by both parties is voter suppression \cite{wang2012politics}. Specifically, due to recent voter identification laws, there is an increased chance of voter suppression \cite{hajnal2017voter}. However, in this work we seek to find instances of voter suppression from an online social media analysis. To our knowledge, this has not been done before.  

\subsection{Political Manipulation}

Social media serve as convenient platforms for people to connect and to exchange ideas. However, social media networks like Twitter and Facebook can be used for malicious purposes \cite{ferrara2015manipulation}. Especially in the context of political discussion, there is a significant risk of mass manipulation of  public opinion. Concerning the ongoing investigation of Russian meddling in the 2016 US Presidential election,  \citet{Badawy2018} studied political manipulation by analyzing the released Russian troll accounts on Twitter. After using label propagation to assign political ideology, they found that Conservatives retweeted Russian trolls over 30 times more than Liberals and produced 36 times more tweets. More recently, \citet{stella2018bots} highlighted how bots can play significant roles in targeting influential humans to manipulate online discussion thus increasing in-fighting. Especially for the spread of fake news, various studies showed how political leaning \cite{allcott2017social}, age \cite{Grinberg2019}, and education \cite{scheufele2019science} can greatly affect fake news spread, alongside with other mechanisms that leverage emotions \cite{ferrara2015quantifying, ferrara2015measuring} and cognitive limits \cite{PENNYCOOK2018, pennycook2019cognitive}. Additionally, \citet{dutt2018senator} showed how foreign actors can more so than just backing one candidate or the other, often manipulate social media for the purpose of sowing discord. 

\subsection{Bias}
Besides manipulation, other potential problems may affect data originating from online social systems. Selection bias is one such example. Concisely, this bias yields a statistically non-representative sample of the true population. A main concern outlined by \citet{ruths2014social}, and to a lesser degree by \citet{malik2015population}, is that social media samples are not representative of the whole voting population because users self-select to participate on the platform and in specific online discussions.
Each social media platform has its own set of biases. 
\citet{mislove2011understanding} looked specifically at the Twitter population from a location, gender, and ethnicity viewpoint. 
From a location perspective, they found underrepresented counties in the Mid-West and over-represented counties in highly dense urban areas \cite{mislove2011understanding}. 
Biases in the representation of gender \cite{pitoura2018measuring}, ethnicity \cite{chang2010epluribus}, and other sources of distortions \cite{culotta2014reducing}  can also potentially affect the inference of political ideology. 

%% file: src/data.tex
\section{Data}

In this study, we examine different data sources to investigate and explore the risk of using social media in the context of political election manipulation. 

We used Twitter as a sensor to estimate the expected amount of votes generated by each state. For this purpose, we carried out two data collections. In the first one, we gathered the tweets with the hashtag \#ivoted on election day. The second collection aimed to enlarge the spectrum to a longer period of time exploiting a variety of general keywords, related to the midterm election, to collect the tweets.
As a basis for comparison, we employ two external sources. 
The United States Election Project is used to unveil the amount of voters in each state, while Gallup to have an estimate of the political polarization both at the country level and at the state level.
By means of these three data sources, we assembled five data sets (DS1-DS5), which will be analyzed in turn in the following subsections.

\subsubsection{DS1: \#ivoted Dataset} 
\begin{figure}[t]
    \centering
    \includegraphics[width=.5\columnwidth]{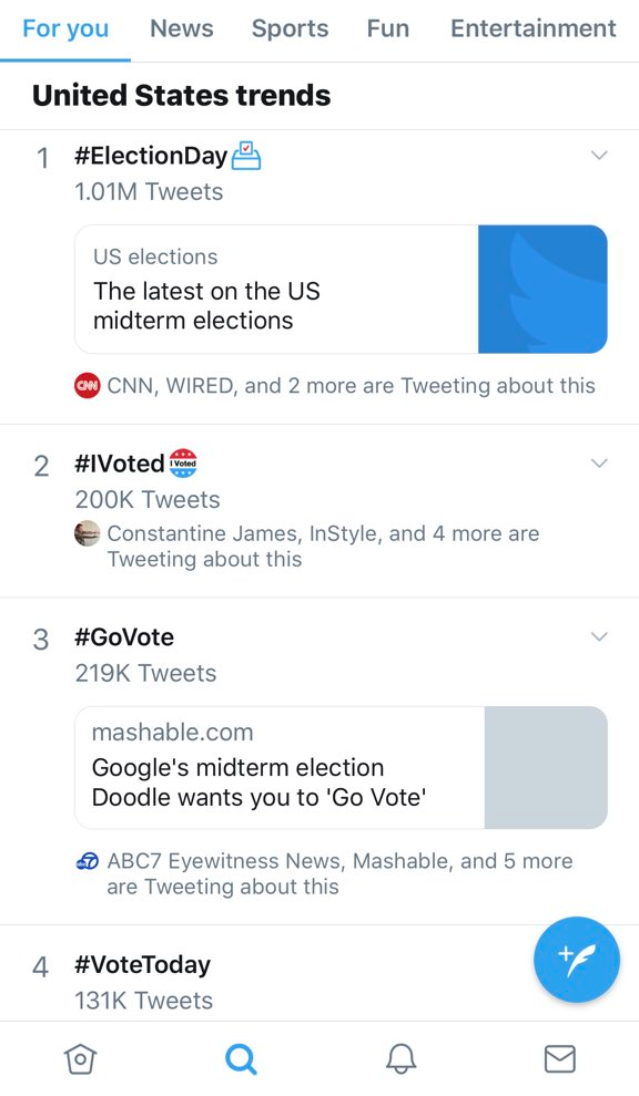}
    \caption{Screen shot of the United States trends on election day showing the \#ivoted hashtag trending with 200K tweets.}
    \label{fig:screenshot}
\vspace{-.5cm}
\end{figure}

The \#ivoted Dataset (DS1) gathers the tweets with the hashtag \#ivoted generated on the day of the election, November 6, 2018.
It should be noticed that \#ivoted was promoted by Twitter and Instagram---which typically affects the hashtag spread \cite{ferrara2016detection, varol2017early}---to encourage citizens to participate in the midterm elections and increase the voter turnout.
We used the Python module \textit{Twyton} to collect tweets through the Twitter Streaming API\footnote{Please note that we utilize the same approach for every Twitter data collection discussed in this work.} during  election day.
The data collection time window ranged from 6 a.m. EST on November 6 (when the first polling station opened) to 1 a.m. HST on November 7 (2 hours after the last polling station closed).
Overall, we collected 249,106 tweets. 
As a sanity check, we queried the \textit{OSoMe API} provided by Indiana University~\cite{davis2016osome}. OSoMe tracks the Twitter Decahose, a pseudo-random 10\% sample of the stream, and therefore can provide an estimate of the total volume: \textit{OSoME} contains 29.7K tweets with the  \#ivoted hashtag posted by 27.2K users---it is worth noting that trending topics are typically slightly over-represented in the Twitter Decahose \cite{davis2016osome, morstatter2013sample}---by extrapolation, this would suggest an estimated upper bound of the total volume at around 300K tweets. In addition, on election day, Twitter reported that the hashtag \#ivoted was trending with over 200K tweets (cf. Fig.~\ref{fig:screenshot}).
Having collected 249K such tweets, we can conclude that we have at our disposal a nearly complete \#ivoted sample dataset.

\subsubsection{DS2 \& DS3: General Midterm Dataset}
In the General Midterm Dataset, we collect tweets on a broader set of keywords.
Further, we consider two different time windows for the data collection.
The rationale behind these choices is to evaluate the sensitivity of our study against a different, but correlated, set of data.
In other words, the main purpose is to detect whether any divergence arose with the \#ivoted Dataset analysis or, on the other hand, to inspect the consistency of the results in different settings.

Tweets were collected by using the following keywords as a filter: \textit{2018midtermelections}, \textit{2018midterms}, \textit{elections}, \textit{midterm}, and \textit{midtermelections}.
We distinguish two data sets according to their temporal extent. In DS2, we consider only tweets generated on the election day with exactly the same time window used for DS1. 
The third data set (DS3) provides a view of the political discussion from a wide-angle lens. It includes tweets from the month prior (October 6, 2018) to two weeks after (November 19, 2018) the day of the election.
We kept the collection running after the election day as several races remained unresolved.
As a result, DS3 consists of 2.7 million tweets, whose 
IDs are publicly available for download.\footnote{\url{https://github.com/A-Deb/midterms}}

\subsubsection{DS4: Actual Voting Data}
The first external data source used as a basis of comparison is made available by the United States Election Project. They report on their website\footnote{\url{http://www.electproject.org/2018g}} the expected voter turnout per  state, along with the (official or certified) information source and other statistics about voters.
The data (DS4) we use in this work was assessed on November 18, 2018, and reflects a voter turnout of 116,241,100 citizens, which is aligned with other reported counts. 

\subsubsection{DS5: Party Affiliation Data}
To have an assessment of the political party affiliation across the country, we make use of an evaluation provided by Gallup, through the Gallup Daily tracking survey, a system which \textit{continuously monitors Americans' attitudes and behaviors}.\footnote{https://www.gallup.com/174155/gallup-daily-tracking-methodology.aspx}
The data set (DS5), collected on January 22, 2019, depicts the political leaning over a sample size of 180,106 citizens.
In particular, the data shows the percentage of Democratic and Republican population in each state and over the entire country.
Gallup's evaluation shows that, at the national level, there exists a democratic advantage (7\%), as
45\% of the population is assessed as democratic leaning while 38\% is estimated as republican.

\subsection{Data Pre-processing}
Data pre-processing involved only Twitter data sets and consisted of three main steps.
First, we removed any duplicate tweet, which may have been captured by accidental duplicate  queries to the Twitter API.
Then, we excluded from our analysis all the tweets not written in English language.
Despite the majority of the tweets were in English, and to a very lesser degree in Spanish (3,177 tweets), we identified about 59 languages in the collected tweets. 
Finally, we inspected tweets from other countries and removed them as they were out of the context of this study.
In particular, we filtered out tweets related to the Cameroon election (October 7, 2018), to the Democratic Republic of the Congo presidential election (December 23, 2018), to the Biafra call for Independence (\#biafra, \#IPOB), to democracy in Kenya (\#democracyKE), to the two major political parties in India (BJP and UPA), and to college midterm exams.

Overall, we count for almost 3 millions tweets distributed over the three Twitter data sets (DS1-DS3).
In Table \ref{tab:datasets_stats}, we report some aggregate statistics. It should be noticed that the number of authors is lower than the number of users, which in turn also includes accounts that got a retweet (or reply) of a tweet that was not captured in our collection and, thus, they do not appear as \textit{authors}.

\begin{table}[t]
\centering \small
\caption{Datasets Statistics}
\label{tab:datasets_stats}
\begin{tabular}{|l|c|c|c|}
\hline
    Statistic & DS1   & DS2 & DS3   \\ 
    
\hline
\# of Tweets	&	90,763	&	20,450	&	452,288	\\
\# of Retweets	&	146,546	&	54,866	&	1,869,313	\\
\# of Replies	&	11,797	&	6,730	&	267,973	\\
\# of Authors	&	174,854	&	72,022	&	977,996	\\
\# of Users	&	178,503	&	77,749	&	997,406	\\
\hline
\end{tabular} 
\vspace{-.5cm}
\end{table}

%% file: src/methodology.tex
\section{Methodology}

\subsection{State Identification} \label{state}
The usage of geo-tagged tweets to assign a state to each user has been shown to not be effective, being the fraction of geo-tagged tweets around 0.5\% \cite{cheng2010you}. 
The location of the data is of utmost importance, especially at the state and local level. However, less than 1\% of the collected tweets have been geo-tagged. Nevertheless, we aim to map as many users as possible to a US state, to conduct a state by state comparison. For this purpose, we leveraged tweet metadata, which may include the self-reported user profile location.
The \textit{location} entry is a user-generated string (up to 100 characters), and it is pulled from the user profile metadata for every tweet. 
From this field, we first search for the two-letter capitalized state codes, followed by the full name of the state. Our analysis does not include Washington, D.C., so we have to ensure anything initially labeled \textit{Washington} does not include any variant of \textit{DC}. Using this string-search method,
we managed to assign a state to approximately 50\% of the tweets and 30\% of the users. Some users had multiple states over their tweet history, thus, we only used the most common reported state. A few users often switched their location from a state name to something else: for example, one user went from \textit{New York, NY} to \textit{Vote Blue!}---for such users, we kept the valid state location.

\subsection{Bot Detection}
Bot detection has received ample attention \cite{ferrara2016rise} and increasingly sophisticated techniques keep emerging \cite{kudugunta2018deep}.
In this study, we restrict our bot detection analysis to the use of the widely popular \textit{Botometer},\footnote{\url{https://botometer.iuni.iu.edu/}} developed by Indiana University. The underpinnings of the system were first published in \cite{davis2016botornot,varol2017online} and further revised in \cite{yang2019arming}. 
Botometer is based on an ensemble classifier \cite{breiman2001random} fed by over 1,000 features related to the Twitter account under analysis and extracted through the Twitter API. Botometer aims to provide an indicator, namely \textit{bot score}, that is used to classify an account either as a bot or as a human. The lower the bot score, the higher the probability that the user is not an automated and/or controlled account. In this study we use version v3 of Botometer, which brings some innovations and important detailed in \cite{yang2019arming}---e.g., the bot scores are now rescaled and not centered around 0.5 anymore.

\begin{figure}[t] 
\centering
  \includegraphics[width=\columnwidth]{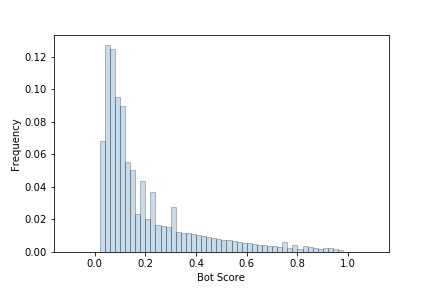}
  \caption{Bot Score Distribution }
  \label{fig:bot_score_dist}
\vspace{-.5cm}
\end{figure}

In Figure \ref{fig:bot_score_dist}, we depict the bot score distribution of the 1,131,540 distinct users in our datasets. 
The distribution exhibits a right skew: most of the probability mass is in the range [0, 0.2] and some peaks can be noticed around 0.3. 
Prior studies used the 0.5 threshold to separate humans from bots. However, according to the re-calibration introduced in the latest version of Botometer \cite{yang2019arming}, along with the emergence of increasingly more sophisticated bots, we here lower the bot score threshold to 0.3 (i.e., a user is labeled as a bot if the bot score is above 0.3). This threshold corresponds to the same level of sensitivity setting of 0.5 in prior versions of Botometer (cf. Fig 5 in \cite{yang2019arming}).
In both DS1 and DS3, 21.1\% of the users have been classified as bots, while in DS2 the percentage achieves the 22.9\% of the users. Finally, 19.5\% of the 295,352 users for which a State was identified have been scored as bots.

Overall, Botometer did not return a score for 42,904 accounts, which corresponds to 3.8\% of the users.
To further examine this subset of users, we make use of the Twitter API.
Interestingly, 99\% of these accounts were suspended by Twitter, whereas the remaining 1\% were protected (by privacy settings).
For the users with an assigned location, only 1,033 accounts did not get a Botometer score. For those users, we assume that the accounts suspended (1,019) are bots and the private accounts (14) are humans.

    

\subsection{Statistical Vote Comparison} 

Once the states have been identified and the bots detected, we  compared the distribution of our various Twitter datasets (DS1, DS2, and DS3) with our control data in DS4 and DS5. To do this, we start by counting the number of tweets per state and dividing it by the total number of tweets across all states. We denote this fractional share in terms of tweets as \textit{State Tweet Rate (STR)}, for each state $i$ as 

\begin{equation}
 STR(i)=\frac{ \mbox{no. tweets from State i }}{\sum_{j}^{50}{ \mbox{no. tweets from State j}}}
 \end{equation}

For the actual voter data (DS4), we perform a similar metric to determine the \textit{State Vote Rate (SVR)} of each state $i$ as

\begin{equation}
 SVR(i)=\frac{\mbox{no. votes from State i}}{\sum_{j}^{50}{\mbox{no. votes from State j}}}
 \end{equation}

We then calculate the difference $\delta(i)$ for each state $i$. Here it is important to note that any positive value indicates more tweets than votes, as  a percentage, and vice versa:
\begin{equation}
 \delta(i)=STR(i)-SVR(i)
 \end{equation}

Lastly, we convert the difference into standard deviations $s(i)$ (stdevs) by dividing $\delta(i)$ by the standard deviation of all differences:
\begin{equation}
 s(i)=\frac{\delta(i)}{\sqrt{\frac{\sum{(\delta(i)-\overline{\delta})}}{50}}}
 \end{equation}

being $\overline{\delta}$ the average difference over all states. We then inspect the results for any anomalous state $i$ whose standard deviation $|s(i)|\geq 2$. States beyond two standard deviations are worth further inspection.

\subsection{Political Ideology Inference}
We classify users by their ideology based on the political leaning of the media outlets they share. 
 We use lists of partisan media outlets compiled by third-party organizations, such as AllSides\footnote{\url{https://www.allsides.com/media-bias/media-bias-ratings}} and Media Bias/Fact Check.\footnote{\url{https://mediabiasfactcheck.com/}} We combine liberal and liberal-center media outlets into one list and conservative and conservative-center into another. The combined list includes 641 liberal and 398 conservative outlets. However, in order to cross reference these media URLs with the URLs in the Twitter dataset, we need to get the expanded URLs for most of the links in the dataset, since most of them are shortened. As this process is quite time-consuming, we get the top 5,000 URLs by popularity and then retrieve the long version for those. These top 5,000 URLs account for more than 254K, or more than 1/3 of all the URLs in the dataset. 
After cross-referencing the 5,000 long URLs with the media URLs, we observe that 32,115  tweets in the dataset contain a URL that points to one of the liberal media outlets and 25,273 tweets with a URL pointing to one of the conservative media outlets. 
%
%
%
We use a polarity rule to label Twitter users as liberal or conservative depending on the number of tweets they produce with links to liberal or conservative sources. In other words, if a user has more tweets with URLs to liberal sources, he/she is labeled as liberal and vice versa. Although the overwhelming majority of users include URLs that are either liberal or conservative, we remove any user that has equal number of tweets from each side. Our final set of labeled users includes 38,920 users.


To classify the remaining accounts as liberal or conservative, we use label propagation, similar to prior work \cite{Badawy2018}. 
For this purpose, we construct a retweet network, containing nodes (Twitter users) with a direct link between them if one user retweet a post of another.  
To validate results of the label propagation algorithm, we apply stratified cross (5-fold) validation to a set of more than 38,920 seeds. We train the algorithm on 4/5 of the seed list and see how it performs on the remaining 1/5. Both precision and recall scores are around 0.89. Since we combine liberal and liberal-center into one list (same for conservatives), we can see that the algorithm is not only labeling the far liberal or conservative correctly, which is a relatively easier task, but it is performing well on the liberal/conservative center as well. 
Overall, we find that the liberal users population is almost three times larger the conservative counterpart (73\% vs. 27\%).

%% file: src/results.tex
\begin{figure}[t]
  \centering
  \includegraphics[width=.87\columnwidth]{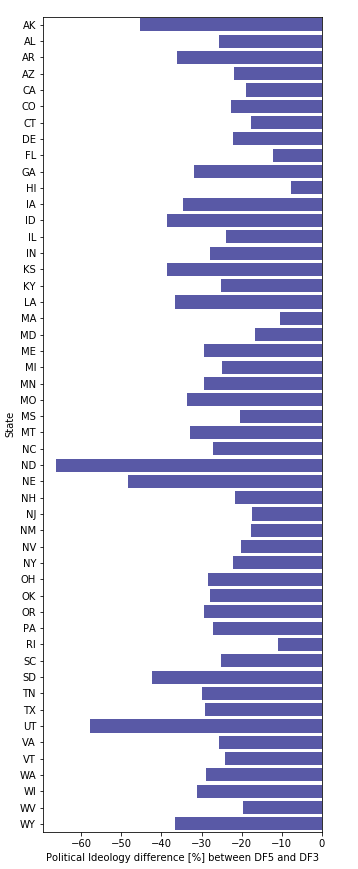}
\caption{Political ideology difference, in terms of percentage of liberals vs. conservatives, between DS5 and DS3}
\label{fig:pol_diff}   \vspace{-.75cm}
\end{figure}

\section{Results}

\subsection{\#ivoted (DS1) Statistical Analysis}

There were 249,106 tweets in the \#ivoted data set, of those we could map a state location for 78,162 unique authors. Once we remove the 15,856 bots (using a bot threshold score of 0.3), we have 62,306 remaining authors of tweets and retweets. After applying the method described in Statistical Vote Comparison section, we see that three states show an anomalous behavior from the remaining 47 states. 
Figure \ref{fig:ivoted1} shows how New York is 5.8 standard deviations greater than the mean difference between the \#ivoted percentage and the actual voting percentage. Furthermore, both California and Texas have a stdev 2.2 greater than the mean. This would lead to believe that if there was voter suppression, it would most likely be in these three states, as they exhibit significantly more self-reported voting tweets than vote counts. 

However, since our data set has both tweets and retweets, to check the sensitivity of our findings, we repeated our analysis without the retweets. Once removed, the 34,754 remaining tweets, again without bots, we noticed something interesting. 
Not only did Texas drop from 2.2 stdevs to 0.4 stdevs, but New York increased from 5.8 stdevs to 6.3 stdevs.
This highlights the sensitivity our this type of analysis to  location-specific factors such as state, and information dynamic factors such as retweet filtering.
Further inspection showed that 62.2\% of the tweet activity in Texas (in the \#ivoted data set) was based on retweets, highlighting how this class of tweet can produce different results for some populations, and similar ones for others, since the average across the states stayed at 0 (e.g., see Figure \ref{fig:ivoted2}).

\subsection{General Midterm (DS2\&DS3) Statistical Analysis}
We carried out the same analysis against the
general keywords data set both on election day (DS2) and for a month before to two weeks after the election (DS3).

\begin{figure*}
        \centering
        \begin{subfigure}[b]{0.475\textwidth}
            \centering
            \includegraphics[width=\textwidth]{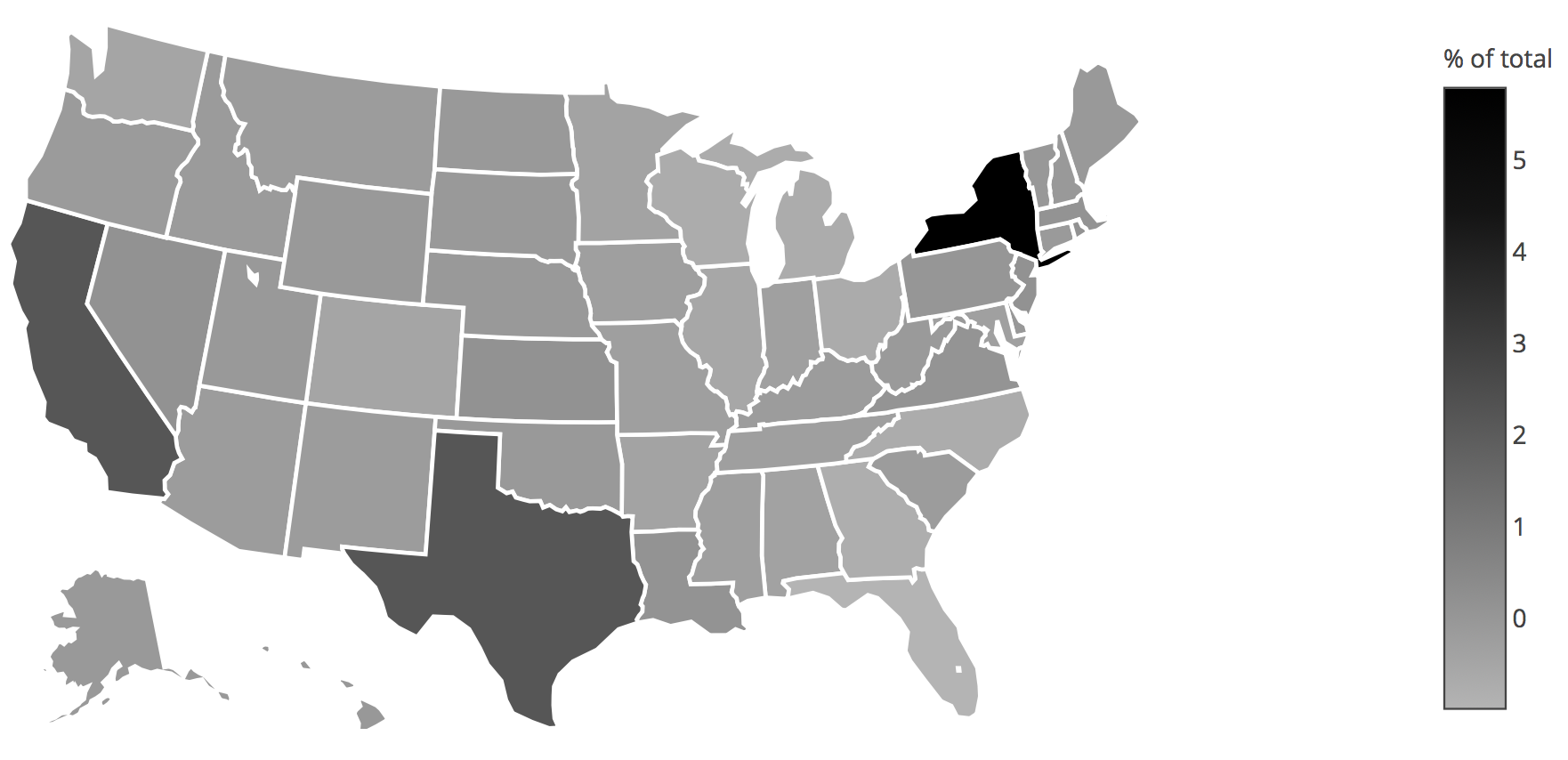}
            \caption[]%
            {{\small \#ivoted vs. Actual Votes}}    
            \label{fig:ivoted1}
        \end{subfigure}
        \hfill
        \begin{subfigure}[b]{0.475\textwidth}  
            \centering 
            \includegraphics[width=\textwidth]{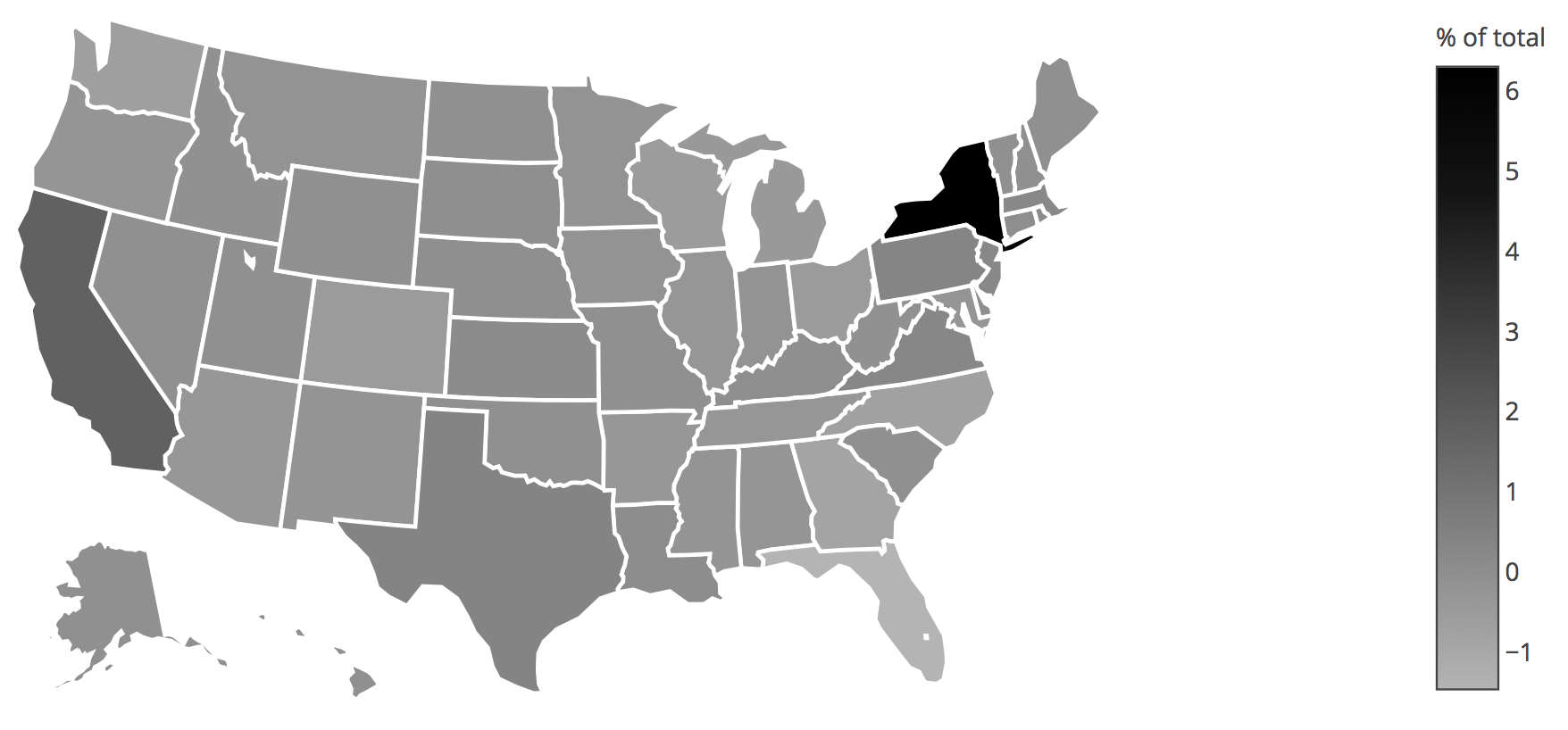}
            \caption[]%
            {{\small \#ivoted (w/o RTs) vs. Actual Votes}}    
            \label{fig:ivoted2}
        \end{subfigure} \\
        \begin{subfigure}[b]{0.48\textwidth}   
            \centering 
            \includegraphics[width=\textwidth]{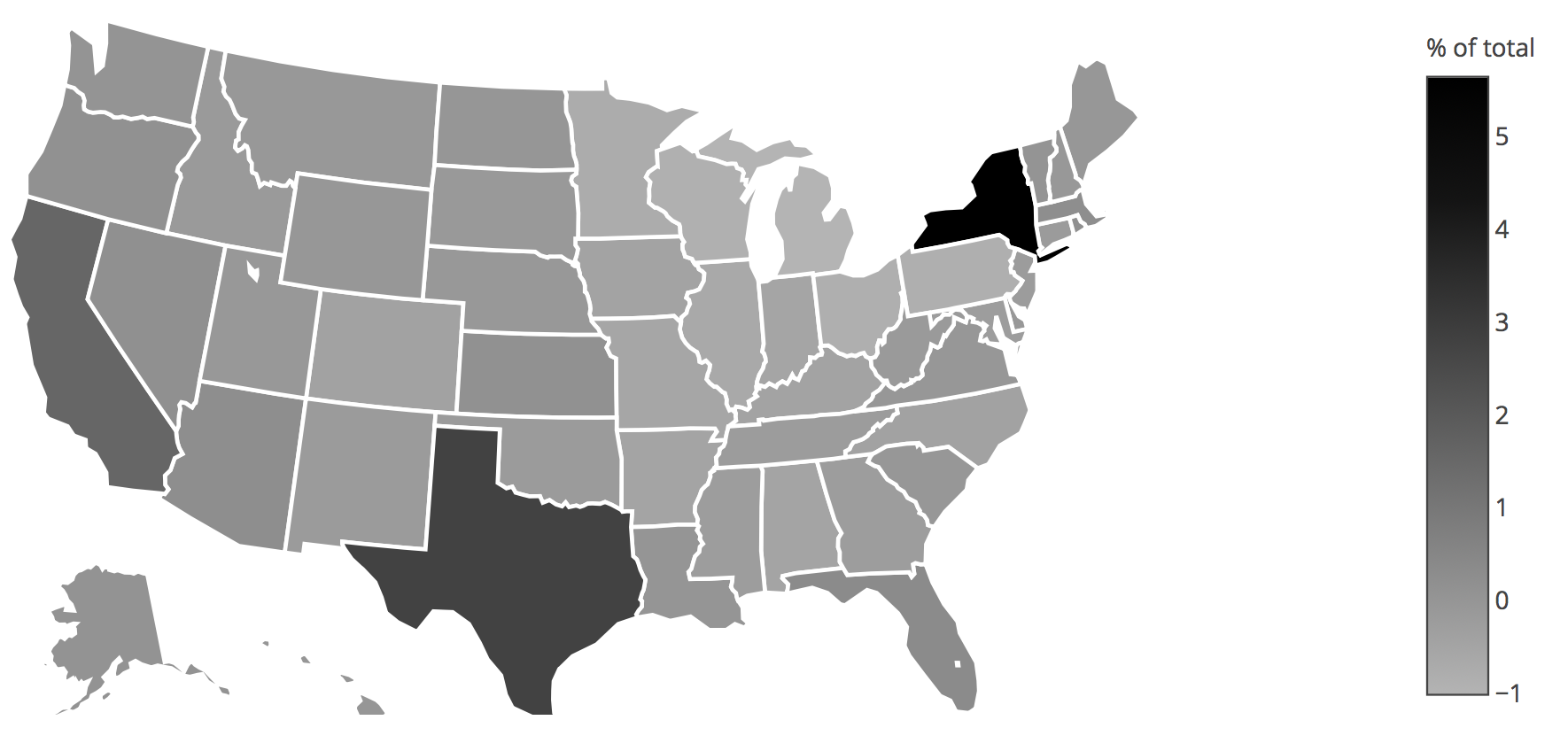}
            \caption[]%
            {{\small General (election) vs. Actual Votes}}    
            \label{fig:general1}
        \end{subfigure}
        \quad
        \begin{subfigure}[b]{0.48\textwidth}   
            \centering 
            \includegraphics[width=\textwidth]{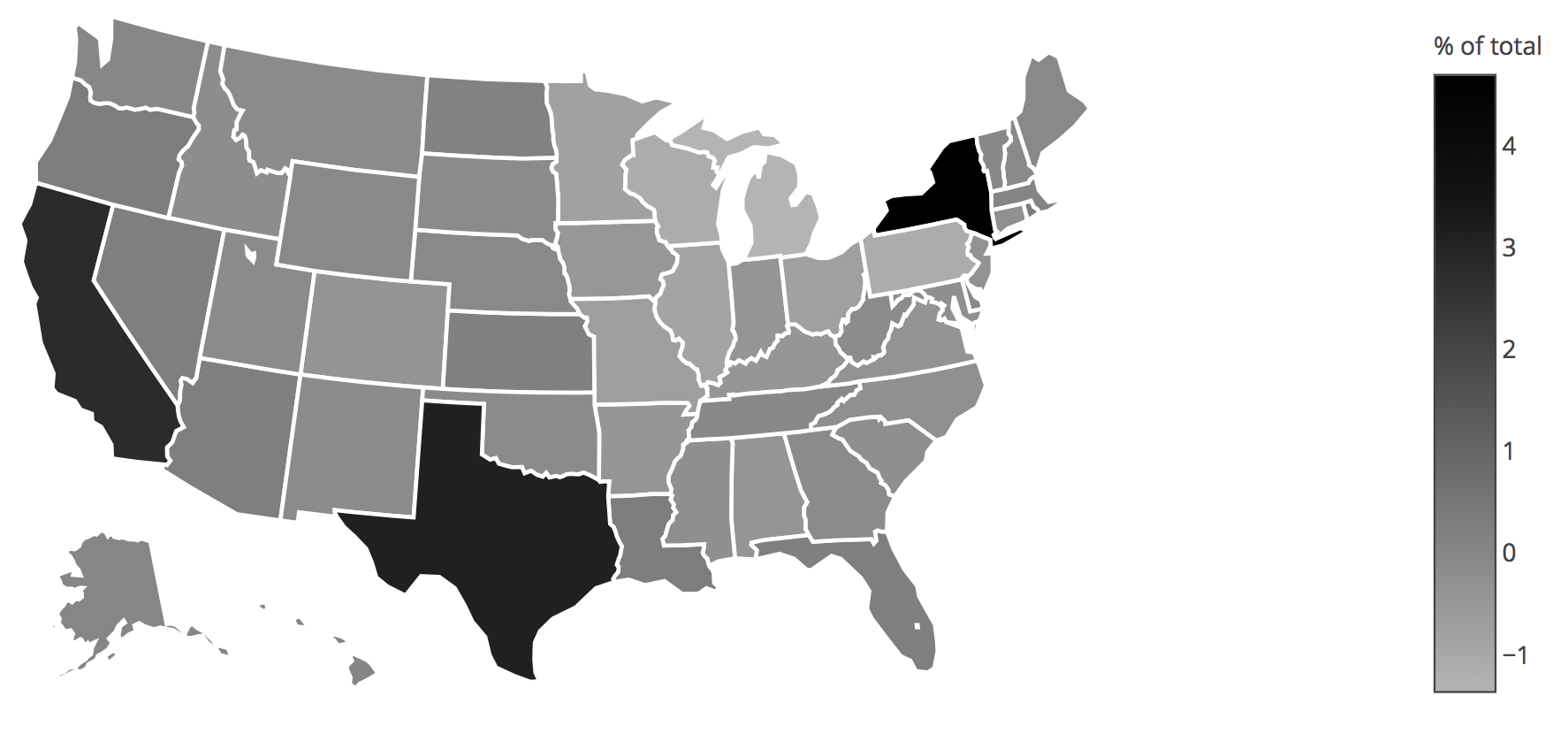}
            \caption[]%
            {{\small General (overall) vs. Actual Votes}}    
            \label{fig:general2}
        \end{subfigure}
        \caption[]
        {\small Various datasets versus Actual Votes (DS4) all without bots} 
        \label{fig:maps}
    \vspace{-.5cm}
    \end{figure*}

In DS2, we have 72,022 users, from which we filtered out 16,859 bots (using a bot threshold of 0.3).
From the remaining 55,163 authors, we were able to map a state for 26,081 users. 
Performing the same comparative analysis from before, we found the same anomalies in the same three states: CA (1.6 stdev), TX (2.8 stdev), and NY (5.6 stdev). Visually, this can be appreciated in Figure \ref{fig:general1}. 
Expanding the analysis to DS3, we removed 206,831 users, as classified as bots, from the set of 977,966 authors.
This left us with 771,135 users from which we could identify a state for 295,705 of them.
The statistical analysis revealed the same outliers also in this data set: CA (2.8 stdev), TX (3.1 stdev), and NY (4.7 stdev), as can been seen in Figure \ref{fig:general2}.

\subsection{Bot Sensitivity}
Next, we investigate whether discarding malicious accounts, such as social bots, from the set of users may have affected the findings above.
Table \ref{tab:bot_per_state} shows the number (and percentage) of bots and humans per state in DS3. The list of states is sorted (in descending order) according to the percentage of bots, while the horizontal line separates the states with a bots percentage above and below the average (20.3\%).
Note in particular that all the three outliers (in bold) have values below the average.
However, the distribution of bot prevalence per state varies greatly and it should be analyzed taking into account both the state population size and the number of Twitter users per state.
Highly populated states like California, Texas, and New York, have large sheer numbers of bots but low proportional bot percentage. This should be taken into account when drawing conclusions from this analysis. On the other side, this topic opens the way to further discussions about bots association with a given state. 
One could make the argument that if the account was identified as a bot, there is no point to assigning it to a state. However, the fact that automated accounts declare a location in their profile can be viewed as a malicious strategy to embed in the social system thus, it should be prudently examined.

For these reasons, we repeated our analysis including social bots in the users set. Results with or without bots are substantially unchanged.
In the interest of space, we do not duplicate the maps shown in Figure \ref{fig:maps}, but the same anomalies are revealed if bots are retained. It should be noticed that also for the \#ivoted dataset (DS1), the percentage of bots in the three outlier states are below the average (21.0\%), NY (16.0\%), CA (19.4\%) and TX (20.2\%), respectively. 


\subsection{Political Ideology Analysis}

Next we examine what topics talk about and how they address politically charged topics. Table \ref{tab:tophashtags} shows the top 10 hashtags discussed respectively by humans and bots, for both liberal and conservative ideologies. The hashtags have been colored to show the common topics between bots and humans for each political wing. The amount of overlap between bots and humans hashtags is noticeable.
This is likely the reason why the removal of bots from the analyzed accounts did not have any significant impact on our outcome. To carefully interpret this table, it should be noticed that the liberal group is almost three times larger than the conservative one, as we stated in \textit{Political Ideology} section. 

Additionally, we took our political ideology labels by state and compared with DS5, the Gallup poll survey. As mentioned before, the political ideology inference assigned 73\% liberal labels and 27\% conservative labels to the nation at a whole. That compares with Gallup reporting of 45\% to 38\% for the Nation as a whole. At the state level, we ran a comparison to see the difference in our assessment of political leaning of a state versus Gallup's. For example, Alabama is 35\% liberal and 50\% conservative, according to Gallup, giving the state a marked Republican advantage.  However, in Twitter we observed 42\% Liberal and 31\% Conservative user labels, which may suggest the opposite trend. Figure \ref{fig:pol_diff} shows the difference between the Gallup poll and our analysis. For Alabama going from a Republican advantage of 15\% (Gallup) to a Democratic advantage of 11\% (Twitter) would imply a shift of 26 percent points toward the liberal side. Overall, every state showed movement toward the left, as low as a few percent points and as high as over 60\% difference. This corroborates the suspect that left-leaning users are over-represented in our data.

\begin{table}[t]
    \centering\small
\begin{tabular}{ |c|c|c| }
\hline
\multicolumn{3}{ |c| }{Top 10 Hashtags} \\
\hline
 & Liberal & Conservative \\ \hline
\multirow{10}{*}{Bots} & \textcolor{blue}{\#BlueWave} & \textcolor{red}{\#BrowardCounty} \\
 & \textcolor{blue}{\#VoteBlue} & \textcolor{red}{\#MAGA} \\
 & \#MAGA & \textcolor{red}{\#Broward} \\
 & \textcolor{blue}{\#NovemberisComing} & \textcolor{red}{\#RedWave} \\
 & \#TheResistance & \#VoteRedToSaveAmerica \\
 & \#Democrats & \textcolor{red}{\#StopTheSteal} \\
 & \textcolor{blue}{\#Trump} & \#VoteRed \\
 & \textcolor{blue}{\#vote} & \#Democrats \\
 & \#Florida & \textcolor{red}{\#Redwavepolls} \\
 & \#GOTV & \#WednesdayWisdom \\ \hline
\multirow{10}{*}{Humans} & \#NovemberisComing & \textcolor{red}{\#BrowardCounty} \\
 & \textcolor{blue}{\#VoteBlue} & \textcolor{red}{\#Broward} \\
 & \textcolor{blue}{\#BlueWave} & \textcolor{red}{\#MAGA} \\
 & \textcolor{blue}{\#vote} & \#IranRegime \\
 & \#txlege & \#Tehran \\
 & \#electionday & \textcolor{red}{\#StopTheSteal} \\
 & \#Russia & \textcolor{red}{\#RedWave} \\
 & \#unhackthevote & \#PalmBeachCounty \\
 & \#AMJoy & \textcolor{red}{\#Redwavepolls} \\
 & \textcolor{blue}{\#Trump} & \#Florida \\ \hline
\end{tabular}
    \caption{Top 10 hashtags: liberals, conservatives, humans,  bots}
    \label{tab:tophashtags}
\vspace{-.5cm}
\end{table}

\subsection{Voting Issues}
\begin{figure}[t]
  \centering\includegraphics[width=.75\columnwidth]{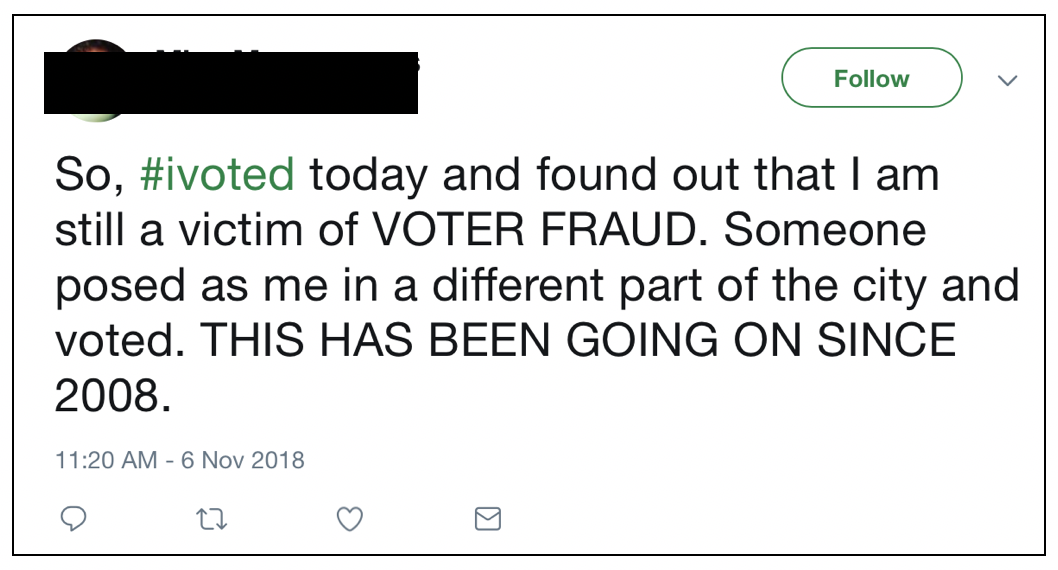}
\caption{\#ivoted tweet from New York}
\label{fig:tweet1}   
\vspace{-.5cm}
\end{figure}

New York was the state that exhibited the strongest statistical anomaly. Thus, we conducted a manual inspection reading all tweets originating from there. We found no red flags, but we isolated a few tweets of interest. The first one is in Figure \ref{fig:tweet1} and it is from a user who was classified as a human and from inspection of the account shown to live in New York. The user mentions some important issues: at 11:20 am on the day of the election, they found out they are the victim of voter fraud. There is no information to suggests this was resolved in any meaningful way or if the accusation is substantiated. 

\begin{figure}[t]
  \centering\includegraphics[width=.75\columnwidth]{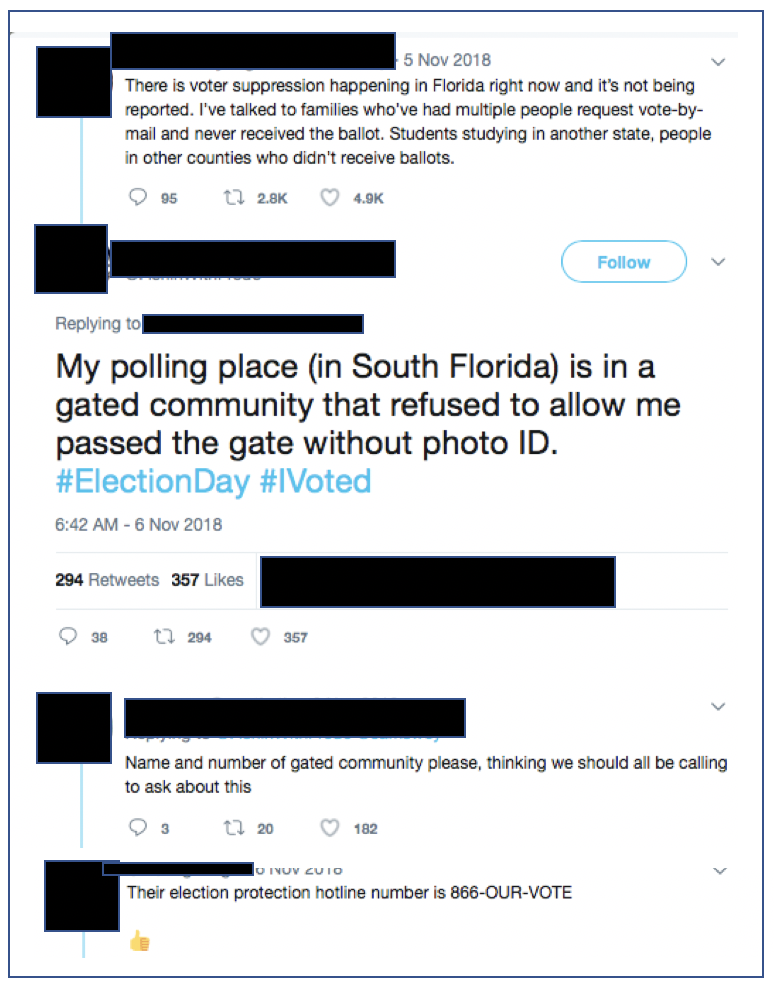}
\caption{\#ivoted tweet from Florida}
\label{fig:tweet2}   
\vspace{-.5cm}
\end{figure}

A second example of potential voter issue was found after a manual inspection of the tweets in New York. The tweet thread in Figure \ref{fig:tweet2} is heavily redacted, but it shows an ongoing conversation through replies and it shows multiple people presenting multiple sides. The original tweet was actually posted on 5 November, 2018 and by the time of our viewing had received a significant number of retweets. It is from this original tweet that we see a reply where the user is complaining that they can not get to the voting booth without a photo ID. User 3 then asks for the name and number of the community and then User 4 provides an election hotline number. This indicates that many people today are willing to speculate on Twitter, but nothing seems to indicate that they also were going to the official Department of Justice website to file a complaint.

From our inspection other tweets that are noteworthy include:
\begin{enumerate}
    \item "First time voter in my family registered over a month ago on DMV website online not realizing it's not automated\ldots she could not vote. Not right."
    \item "More voter fraud in Ohio. Why is it that all the errors are always the Democrats?? Because the only way they can win is if they cheat!! This madness needs to stop." 
    \item What we did see in our Twitter collection is early skepticism that there would be false claims of voter fraud. A user tweeted "a little over 24 hours from now the Racist in Chief will start Tweeting about rigged elections, voter fraud and illegal aliens voting en mass...".
    \item Shortly afterwards, many people started to retweet a user that stated "Massive voter fraud in Texas Georgia Florida and others" and also indicating that MSM (main stream media) are putting out fake polls. The Washington Post $@$washingtonpost tweeted "without evidence, Trump and Sessions warn of voter fraud" which was retweeted throughout election day. 
    \item There was a user who tweeted about voting machine malfunctions which mapped to a story/blog from the Atlanta Journal Constitution (\url{https://t.com/riCGdbwQ6R}) about machines being down; people left and were encouraged to come back. There was an offer for casting a paper provisional ballot, but many said they did not trust the paper ballot and wanted to vote on a machine. 
\end{enumerate}

\begin{table}[t] 
\caption{General Midterms DS3: bot and human population by State (sorted by percent-wise bot prevalence). 
\label{tab:bot_per_state}} 
\centering
\begin{tabular}{lrr}
\toprule
State & \# of bots  & \# of humans \\
\midrule
WY    &        97 (27.2\%) &         246 (68.9\%) \\
ID    &       258 (23.8\%) &         791          (73.0\%) \\
ND    &       289 (22.9\%) &         931          (73.9\%)\\
AZ    &      1,514 (22.5\%) &        4,997          (74.2\%)\\
NV    &       711 (22.4\%) &        2,377          (74.8\%) \\
UT    &       420 (22.2\%) &        1,425          (75.4\%) \\
DE    &       170 (22.1\%) &         575          (74.9\%) \\
NM    &       325 (22.1\%) &        1,100          (74.7\%) \\
NH    &       283 (22.1\%) &         968          (75.5\%) \\
RI    &       402 (21.9\%) &        1,382          (75.3\%) \\
WV    &       246 (21.7\%) &         854          (75.2\%) \\
FL    &      4,696 (21.5\%) &       16,583          (75.9\%) \\
MO    &       932 (21.4\%) &        3,336          (76.5\%) \\
AL    &       697 (21.3\%) &        2,466          (75.4\%) \\
TN    &      1,209 (21.3\%) &        4,369          (76.9\%) \\
WI    &       808 (21.3\%) &        2,900          (76.4\%) \\
MT    &       202 (21.0\%) &         730          (75.9\%) \\
CO    &      1,144 (20.9\%) &        4,178          (76.4\%) \\
NJ    &      1,311 (20.8\%) &        4,838          (76.8\%) \\
MS    &       336 (20.6\%) &        1,239          (75.9\%) \\
ME    &       290 (20.6\%) &        1,093          (77.6\%) \\
CT    &       571 (20.4\%) &        2,141          (76.6\%) \\
SC    &       769 (20.3\%) &        2,933          (77.5\%) \\
\hline
OK    &       552 (20.2\%) &        2,098          (76.8\%) \\
KS    &       661 (20.2\%) &        2,526          (77.2\%) \\
GA    &      1,962 (20.2\%) &        7,489          (76.9\%) \\
WA    &      1,561 (19.9\%) &        6,143          (78.2\%) \\
NE    &       323 (19.9\%) &        1,253          (77.1\%) \\
AK    &       160 (19.8\%) &         622          (77.1\%) \\
HI    &       230 (19.8\%) &         895          (77.1\%) \\
PA    &      1,898 (19.7\%) &        7,460          (77.6\%) \\
MI    &      1,441 (19.6\%) &        5,714          (77.7\%) \\
IA    &       414 (19.6\%) &        1,654          (78.2\%) \\
VA    &      1,487 (19.6\%) &        5,931          (78.1\%) \\
MA    &      1,372 (19.4\%) &        5,553          (78.4\%) \\
NC    &      1,685 (19.3\%) &        6,872          (78.5\%) \\
IL    &      1,702 (19.2\%) &        6,926          (78.0\%) \\
IN    &       885 (19.1\%) &        3,593          (77.6\%) \\
AR    &       199 (19.1\%) &         814          (78.0\%) \\
KY    &       548 (19.0\%) &        2,270          (78.9\%) \\
MN    &       866 (19.0\%) &        3,622          (79.6\%) \\
OR    &      1,067 (18.9\%) &        4,416          (78.3\%) \\
\textbf{TX}    &      \textbf{5,550 (18.7\%) }&       \textbf{23,448          (79.1\%) }\\
OH    &      1,722 (18.6\%) &        7,271          (78.7\%) \\
\textbf{CA}    &      \textbf{7,073 (18.2\%)} &       \textbf{30,429          (78.5\%)} \\
VT    &       113 (17.9\%) &         505          (80.2\%) \\
MD    &       887 (17.8\%) &        3,963          (79.7\%) \\
\textbf{NY}    &      \textbf{4,798 (17.5\%) }&       \textbf{21,896          (79.9\%)} \\
LA    &       708 (15.6\%) &        3,708          (81.7\%) \\
SD    &        82 (15.4\%) &         439          (82.4\%) \\
\bottomrule
\end{tabular}
\end{table}

%% file: src/discussion.tex
\section{Discussion \& Recommendations}

Our results have highlighted the challenges of using social media in election manipulation analysis. A superficial interpretation of anomalies in online activity  compared to real world data can lead to misleading or false conclusions. In our case, we wanted to determine the feasibility of using social media as a sensor  to detect election manipulation such as  widespread voter suppression or voter fraud. While we did not find widespread or systematic  manipulation, we learned a few lessons worthy of a discussion:

\begin{itemize}
    \item Data biases of online platforms can drastically affect the feasibility of a study. In our case, we were looking for a representative sample of actual voters who are not bots and whose political ideology and location could be known. Despite troves of data were collected and analyzed, various encountered biases  could not be adjusted for.

    \item The second main issue is consistency in the analysis: the sensitivity to choices made when carrying out data cleaning, parameter settings of inference algorithms, etc.  yield a so-called \textit{garden of forking paths} \cite{gelman2013garden}: some results can significantly vary in function of such choices (for example, location bias and the removal or retention of retweets played a role in determining whether Texas exhibited a statistical anomaly in terms of expected versus cast votes).
    
    \item Political ideologies reported by Gallup significantly vary with respect to that can be inferred on social media. We were unable to determine if this is due to limitations of the employed political inference tool, population biases, or other factors. This is an open problem in social media analysis and a necessary one to tackle before social media can be used to robustly replace polling.

    \item The actual voting numbers reported by official sources correlated very closely to what we inferred from our analysis on Twitter for 47 of 50 states. As such, the approach seemed promising to identify voter suppression or fraud. However, the results show a more complex picture: no evidence of fraud or suppression beyond anecdotal was found in the three anomalous states under scrutiny. 
 Yet, we suggest that prior and during elections there should be an online social media presence for the Department of Justice to engage with people who have a potential voting issue. 
\end{itemize}


%% file: src/conclusion.tex
\section{Conclusion and Future Work}
In this work, we conducted an investigation to analyze social media during the 2018 US Midterm election. In addition to studying bots and the political ideology of users, we studied the correlation between people talking about voting and actual voter data. We then highlighted a few issues that could lead to inaccurate conclusions. 
In particular, removing or retaining the bots didn't change the outcome of our results. This was not the case in prior studies. However, in our case, removing retweets did make a significant difference for one state, Texas, suggesting a dependency, or bias, on location. 

The challenges we faced can all be expanded upon in future work. We only mapped a state to 44.7\% of DS1 and 30.2\% to DS2/DS3. If we can evaluate a user timeline to better recognize what state they may be from that would enhance future location based studies.
Our political ideology inference started with the labeling of 38K users leveraging any link they posted, and then  labels were propagated on the retweet network. We could potentially identify the users with high centrality and evaluate their timeline for party affiliation and approach the inference problem from a different angle. 
We could also focus on separating not just human from bot accounts, but also human from corporate accounts. 
Some of the users  that were classified as human could be operating as part of a collective body, that while not necessarily malicious, may insert an inorganic bias. 

Ultimately, one of the goals of this work was to explore the feasibility of using social media as a sensor to detect possible election manipulation at scale: despite our initial effort did not produce the expected results, we highlighted some useful lessons that will illuminate on future endeavors to use such data for social good.

%% file: main.bbl

\begin{thebibliography}{59}


\ifx \showCODEN    \undefined \def \showCODEN     #1{\unskip}     \fi
\ifx \showDOI      \undefined \def \showDOI       #1{#1}\fi
\ifx \showISBNx    \undefined \def \showISBNx     #1{\unskip}     \fi
\ifx \showISBNxiii \undefined \def \showISBNxiii  #1{\unskip}     \fi
\ifx \showISSN     \undefined \def \showISSN      #1{\unskip}     \fi
\ifx \showLCCN     \undefined \def \showLCCN      #1{\unskip}     \fi
\ifx \shownote     \undefined \def \shownote      #1{#1}          \fi
\ifx \showarticletitle \undefined \def \showarticletitle #1{#1}   \fi
\ifx \showURL      \undefined \def \showURL       {\relax}        \fi
\providecommand\bibfield[2]{#2}
\providecommand\bibinfo[2]{#2}
\providecommand\natexlab[1]{#1}
\providecommand\showeprint[2][]{arXiv:#2}

\bibitem[\protect\citeauthoryear{Allcott and Gentzkow}{Allcott and
  Gentzkow}{2017}]%
        {allcott2017social}
\bibfield{author}{\bibinfo{person}{Hunt Allcott} {and} \bibinfo{person}{Matthew
  Gentzkow}.} \bibinfo{year}{2017}\natexlab{}.
\newblock \showarticletitle{Social media and fake news in the 2016 election}.
\newblock \bibinfo{journal}{\emph{Journal of Economic Perspectives}}
  \bibinfo{volume}{31}, \bibinfo{number}{2} (\bibinfo{year}{2017}),
  \bibinfo{pages}{211--36}.
\newblock


\bibitem[\protect\citeauthoryear{Allem, Ferrara, Uppu, Cruz, and Unger}{Allem
  et~al\mbox{.}}{2017}]%
        {allem2017cigarette}
\bibfield{author}{\bibinfo{person}{Jon-Patrick Allem}, \bibinfo{person}{Emilio
  Ferrara}, \bibinfo{person}{Sree~Priyanka Uppu}, \bibinfo{person}{Tess~Boley
  Cruz}, {and} \bibinfo{person}{Jennifer~B Unger}.}
  \bibinfo{year}{2017}\natexlab{}.
\newblock \showarticletitle{E-cigarette surveillance with social media data:
  social bots, emerging topics, and trends}.
\newblock \bibinfo{journal}{\emph{JMIR public health and surveillance}}
  (\bibinfo{year}{2017}).
\newblock


\bibitem[\protect\citeauthoryear{Azzimonti and Fernandes}{Azzimonti and
  Fernandes}{2018}]%
        {azzimonti2018social}
\bibfield{author}{\bibinfo{person}{Marina Azzimonti} {and}
  \bibinfo{person}{Marcos Fernandes}.} \bibinfo{year}{2018}\natexlab{}.
\newblock \bibinfo{booktitle}{\emph{Social media networks, fake news, and
  polarization}}.
\newblock \bibinfo{type}{{T}echnical {R}eport}. \bibinfo{institution}{National
  Bureau of Economic Research}.
\newblock


\bibitem[\protect\citeauthoryear{Badawy, Ferrara, and Lerman}{Badawy
  et~al\mbox{.}}{2018a}]%
        {Badawy2018}
\bibfield{author}{\bibinfo{person}{Adam Badawy}, \bibinfo{person}{Emilio
  Ferrara}, {and} \bibinfo{person}{Kristina Lerman}.}
  \bibinfo{year}{2018}\natexlab{a}.
\newblock \showarticletitle{Analyzing the Digital Traces of Political
  Manipulation: The 2016 Russian Interference Twitter Campaign}. In
  \bibinfo{booktitle}{\emph{Int. Conference on Advances in Social Networks
  Analysis and Mining}}. \bibinfo{pages}{258--265}.
\newblock


\bibitem[\protect\citeauthoryear{Badawy, Lerman, and Ferrara}{Badawy
  et~al\mbox{.}}{2018b}]%
        {badawy2018falls}
\bibfield{author}{\bibinfo{person}{Adam Badawy}, \bibinfo{person}{Kristina
  Lerman}, {and} \bibinfo{person}{Emilio Ferrara}.}
  \bibinfo{year}{2018}\natexlab{b}.
\newblock \showarticletitle{Who Falls for Online Political Manipulation?}
\newblock \bibinfo{journal}{\emph{arXiv preprint arXiv:1808.03281}}
  (\bibinfo{year}{2018}).
\newblock


\bibitem[\protect\citeauthoryear{Bail, Argyle, Brown, Bumpus, Chen, Hunzaker,
  Lee, Mann, Merhout, and Volfovsky}{Bail et~al\mbox{.}}{2018}]%
        {bail2018exposure}
\bibfield{author}{\bibinfo{person}{Christopher~A Bail}, \bibinfo{person}{Lisa~P
  Argyle}, \bibinfo{person}{Taylor~W Brown}, \bibinfo{person}{John~P Bumpus},
  \bibinfo{person}{Haohan Chen}, \bibinfo{person}{MB~Fallin Hunzaker},
  \bibinfo{person}{Jaemin Lee}, \bibinfo{person}{Marcus Mann},
  \bibinfo{person}{Friedolin Merhout}, {and} \bibinfo{person}{Alexander
  Volfovsky}.} \bibinfo{year}{2018}\natexlab{}.
\newblock \showarticletitle{Exposure to opposing views on social media can
  increase political polarization}.
\newblock \bibinfo{journal}{\emph{PNAS}} \bibinfo{volume}{115},
  \bibinfo{number}{37} (\bibinfo{year}{2018}), \bibinfo{pages}{9216--9221}.
\newblock


\bibitem[\protect\citeauthoryear{Bessi and Ferrara}{Bessi and Ferrara}{2016}]%
        {bessi2016social}
\bibfield{author}{\bibinfo{person}{Alessandro Bessi} {and}
  \bibinfo{person}{Emilio Ferrara}.} \bibinfo{year}{2016}\natexlab{}.
\newblock \showarticletitle{Social bots distort the 2016 US Presidential
  election online discussion}.
\newblock \bibinfo{journal}{\emph{First Monday}} \bibinfo{volume}{21},
  \bibinfo{number}{11} (\bibinfo{year}{2016}).
\newblock


\bibitem[\protect\citeauthoryear{Boichak, Jackson, Hemsley, and
  Tanupabrungsun}{Boichak et~al\mbox{.}}{2018}]%
        {boichak2018automated}
\bibfield{author}{\bibinfo{person}{Olga Boichak}, \bibinfo{person}{Sam
  Jackson}, \bibinfo{person}{Jeff Hemsley}, {and} \bibinfo{person}{Sikana
  Tanupabrungsun}.} \bibinfo{year}{2018}\natexlab{}.
\newblock \showarticletitle{Automated Diffusion? Bots and Their Influence
  During the 2016 US Presidential Election}. In
  \bibinfo{booktitle}{\emph{International Conference on Information}}.
  Springer, \bibinfo{pages}{17--26}.
\newblock


\bibitem[\protect\citeauthoryear{Bovet and Makse}{Bovet and Makse}{2019}]%
        {bovet2019influence}
\bibfield{author}{\bibinfo{person}{Alexandre Bovet} {and}
  \bibinfo{person}{Hern{\'a}n~A Makse}.} \bibinfo{year}{2019}\natexlab{}.
\newblock \showarticletitle{Influence of fake news in Twitter during the 2016
  US presidential election}.
\newblock \bibinfo{journal}{\emph{Nature communications}} \bibinfo{volume}{10},
  \bibinfo{number}{1} (\bibinfo{year}{2019}), \bibinfo{pages}{7}.
\newblock


\bibitem[\protect\citeauthoryear{Breiman}{Breiman}{2001}]%
        {breiman2001random}
\bibfield{author}{\bibinfo{person}{Leo Breiman}.}
  \bibinfo{year}{2001}\natexlab{}.
\newblock \showarticletitle{Random forests}.
\newblock \bibinfo{journal}{\emph{Machine learning}} \bibinfo{volume}{45},
  \bibinfo{number}{1} (\bibinfo{year}{2001}), \bibinfo{pages}{5--32}.
\newblock


\bibitem[\protect\citeauthoryear{Chang, Rosenn, Backstrom, and Marlow}{Chang
  et~al\mbox{.}}{2010}]%
        {chang2010epluribus}
\bibfield{author}{\bibinfo{person}{Jonathan Chang}, \bibinfo{person}{Itamar
  Rosenn}, \bibinfo{person}{Lars Backstrom}, {and} \bibinfo{person}{Cameron
  Marlow}.} \bibinfo{year}{2010}\natexlab{}.
\newblock \showarticletitle{ePluribus: Ethnicity on Social Networks.}
\newblock \bibinfo{journal}{\emph{ICWSM}}  \bibinfo{volume}{10}
  (\bibinfo{year}{2010}), \bibinfo{pages}{18--25}.
\newblock


\bibitem[\protect\citeauthoryear{Cheng, Caverlee, and Lee}{Cheng
  et~al\mbox{.}}{2010}]%
        {cheng2010you}
\bibfield{author}{\bibinfo{person}{Zhiyuan Cheng}, \bibinfo{person}{James
  Caverlee}, {and} \bibinfo{person}{Kyumin Lee}.}
  \bibinfo{year}{2010}\natexlab{}.
\newblock \showarticletitle{You are where you tweet: a content-based approach
  to geo-locating twitter users}. In \bibinfo{booktitle}{\emph{CIKM}}.
  \bibinfo{pages}{759--768}.
\newblock


\bibitem[\protect\citeauthoryear{Culotta}{Culotta}{2014}]%
        {culotta2014reducing}
\bibfield{author}{\bibinfo{person}{Aron Culotta}.}
  \bibinfo{year}{2014}\natexlab{}.
\newblock \showarticletitle{Reducing sampling bias in social media data for
  county health inference}. In \bibinfo{booktitle}{\emph{Joint Statistical
  Meetings Proceedings}}. \bibinfo{pages}{1--12}.
\newblock


\bibitem[\protect\citeauthoryear{Davis, Ciampaglia, Aiello, Chung, Conover,
  Ferrara, Flammini, Fox, Gao, Gon{\c{c}}alves, et~al\mbox{.}}{Davis
  et~al\mbox{.}}{2016a}]%
        {davis2016osome}
\bibfield{author}{\bibinfo{person}{Clayton~A Davis},
  \bibinfo{person}{Giovanni~Luca Ciampaglia}, \bibinfo{person}{Luca~Maria
  Aiello}, \bibinfo{person}{Keychul Chung}, \bibinfo{person}{Michael~D
  Conover}, \bibinfo{person}{Emilio Ferrara}, \bibinfo{person}{Alessandro
  Flammini}, \bibinfo{person}{Geoffrey~C Fox}, \bibinfo{person}{Xiaoming Gao},
  \bibinfo{person}{Bruno Gon{\c{c}}alves}, {et~al\mbox{.}}}
  \bibinfo{year}{2016}\natexlab{a}.
\newblock \showarticletitle{OSoMe: the IUNI observatory on social media}.
\newblock \bibinfo{journal}{\emph{PeerJ Computer Science}}  \bibinfo{volume}{2}
  (\bibinfo{year}{2016}), \bibinfo{pages}{e87}.
\newblock


\bibitem[\protect\citeauthoryear{Davis, Varol, Ferrara, Flammini, and
  Menczer}{Davis et~al\mbox{.}}{2016b}]%
        {davis2016botornot}
\bibfield{author}{\bibinfo{person}{Clayton~Allen Davis}, \bibinfo{person}{Onur
  Varol}, \bibinfo{person}{Emilio Ferrara}, \bibinfo{person}{Alessandro
  Flammini}, {and} \bibinfo{person}{Filippo Menczer}.}
  \bibinfo{year}{2016}\natexlab{b}.
\newblock \showarticletitle{Botornot: A system to evaluate social bots}. In
  \bibinfo{booktitle}{\emph{Proceedings of the 25th International Conference
  Companion on World Wide Web}}.
\newblock


\bibitem[\protect\citeauthoryear{Dutt, Deb, and Ferrara}{Dutt
  et~al\mbox{.}}{2018}]%
        {dutt2018senator}
\bibfield{author}{\bibinfo{person}{Ritam Dutt}, \bibinfo{person}{Ashok Deb},
  {and} \bibinfo{person}{Emilio Ferrara}.} \bibinfo{year}{2018}\natexlab{}.
\newblock \showarticletitle{``Senator, We Sell Ads'': Analysis of the 2016
  Russian Facebook Ads Campaign}. In \bibinfo{booktitle}{\emph{International
  Conference on Intelligent Information Technologies}}. Springer,
  \bibinfo{pages}{151--168}.
\newblock


\bibitem[\protect\citeauthoryear{Ferrara}{Ferrara}{2015}]%
        {ferrara2015manipulation}
\bibfield{author}{\bibinfo{person}{Emilio Ferrara}.}
  \bibinfo{year}{2015}\natexlab{}.
\newblock \showarticletitle{Manipulation and abuse on social media}.
\newblock \bibinfo{journal}{\emph{ACM SIGWEB Newsletter}}
  \bibinfo{number}{Spring} (\bibinfo{year}{2015}), \bibinfo{pages}{4}.
\newblock


\bibitem[\protect\citeauthoryear{Ferrara, Varol, Davis, Menczer, and
  Flammini}{Ferrara et~al\mbox{.}}{2016a}]%
        {ferrara2016rise}
\bibfield{author}{\bibinfo{person}{Emilio Ferrara}, \bibinfo{person}{Onur
  Varol}, \bibinfo{person}{Clayton Davis}, \bibinfo{person}{Filippo Menczer},
  {and} \bibinfo{person}{Alessandro Flammini}.}
  \bibinfo{year}{2016}\natexlab{a}.
\newblock \showarticletitle{The rise of social bots}.
\newblock \bibinfo{journal}{\emph{Commun. ACM}} \bibinfo{volume}{59},
  \bibinfo{number}{7} (\bibinfo{year}{2016}), \bibinfo{pages}{96--104}.
\newblock


\bibitem[\protect\citeauthoryear{Ferrara, Varol, Menczer, and Flammini}{Ferrara
  et~al\mbox{.}}{2016b}]%
        {ferrara2016detection}
\bibfield{author}{\bibinfo{person}{Emilio Ferrara}, \bibinfo{person}{Onur
  Varol}, \bibinfo{person}{Filippo Menczer}, {and} \bibinfo{person}{Alessandro
  Flammini}.} \bibinfo{year}{2016}\natexlab{b}.
\newblock \showarticletitle{Detection of promoted social media campaigns}. In
  \bibinfo{booktitle}{\emph{Tenth International AAAI Conference on Web and
  Social Media}}. \bibinfo{pages}{563--566}.
\newblock


\bibitem[\protect\citeauthoryear{Ferrara and Yang}{Ferrara and Yang}{2015a}]%
        {ferrara2015measuring}
\bibfield{author}{\bibinfo{person}{Emilio Ferrara} {and} \bibinfo{person}{Zeyao
  Yang}.} \bibinfo{year}{2015}\natexlab{a}.
\newblock \showarticletitle{Measuring emotional contagion in social media}.
\newblock \bibinfo{journal}{\emph{PloS one}} \bibinfo{volume}{10},
  \bibinfo{number}{11} (\bibinfo{year}{2015}), \bibinfo{pages}{e0142390}.
\newblock


\bibitem[\protect\citeauthoryear{Ferrara and Yang}{Ferrara and Yang}{2015b}]%
        {ferrara2015quantifying}
\bibfield{author}{\bibinfo{person}{Emilio Ferrara} {and} \bibinfo{person}{Zeyao
  Yang}.} \bibinfo{year}{2015}\natexlab{b}.
\newblock \showarticletitle{Quantifying the effect of sentiment on information
  diffusion in social media}.
\newblock \bibinfo{journal}{\emph{PeerJ Computer Science}}  \bibinfo{volume}{1}
  (\bibinfo{year}{2015}), \bibinfo{pages}{e26}.
\newblock


\bibitem[\protect\citeauthoryear{Gayo-Avello}{Gayo-Avello}{2011}]%
        {gayo2011don}
\bibfield{author}{\bibinfo{person}{Daniel Gayo-Avello}.}
  \bibinfo{year}{2011}\natexlab{}.
\newblock \showarticletitle{Don't turn social media into another'Literary
  Digest'poll}.
\newblock \bibinfo{journal}{\emph{Commun. ACM}} \bibinfo{volume}{54},
  \bibinfo{number}{10} (\bibinfo{year}{2011}), \bibinfo{pages}{121--128}.
\newblock


\bibitem[\protect\citeauthoryear{Gayo-Avello}{Gayo-Avello}{2012a}]%
        {gayo2012wanted}
\bibfield{author}{\bibinfo{person}{Daniel Gayo-Avello}.}
  \bibinfo{year}{2012}\natexlab{a}.
\newblock \showarticletitle{" I Wanted to Predict Elections with Twitter and
  all I got was this Lousy Paper"--A Balanced Survey on Election Prediction
  using Twitter Data}.
\newblock \bibinfo{journal}{\emph{arXiv preprint arXiv:1204.6441}}
  (\bibinfo{year}{2012}).
\newblock


\bibitem[\protect\citeauthoryear{Gayo-Avello}{Gayo-Avello}{2012b}]%
        {gayo2012no}
\bibfield{author}{\bibinfo{person}{Daniel Gayo-Avello}.}
  \bibinfo{year}{2012}\natexlab{b}.
\newblock \showarticletitle{No, you cannot predict elections with Twitter}.
\newblock \bibinfo{journal}{\emph{IEEE Internet Computing}}
  \bibinfo{volume}{16}, \bibinfo{number}{6} (\bibinfo{year}{2012}),
  \bibinfo{pages}{91--94}.
\newblock


\bibitem[\protect\citeauthoryear{Gayo-Avello}{Gayo-Avello}{2013}]%
        {gayo2013meta}
\bibfield{author}{\bibinfo{person}{Daniel Gayo-Avello}.}
  \bibinfo{year}{2013}\natexlab{}.
\newblock \showarticletitle{A meta-analysis of state-of-the-art electoral
  prediction from Twitter data}.
\newblock \bibinfo{journal}{\emph{Social Science Computer Review}}
  \bibinfo{volume}{31}, \bibinfo{number}{6} (\bibinfo{year}{2013}),
  \bibinfo{pages}{649--679}.
\newblock


\bibitem[\protect\citeauthoryear{Gayo~Avello, Metaxas, and
  Mustafaraj}{Gayo~Avello et~al\mbox{.}}{2011}]%
        {gayo2011limits}
\bibfield{author}{\bibinfo{person}{Daniel Gayo~Avello},
  \bibinfo{person}{Panagiotis~T Metaxas}, {and} \bibinfo{person}{Eni
  Mustafaraj}.} \bibinfo{year}{2011}\natexlab{}.
\newblock \showarticletitle{Limits of electoral predictions using Twitter}. In
  \bibinfo{booktitle}{\emph{ICWSM}}.
\newblock


\bibitem[\protect\citeauthoryear{Gelman and Loken}{Gelman and Loken}{2013}]%
        {gelman2013garden}
\bibfield{author}{\bibinfo{person}{Andrew Gelman} {and} \bibinfo{person}{Eric
  Loken}.} \bibinfo{year}{2013}\natexlab{}.
\newblock \showarticletitle{The garden of forking paths: Why multiple
  comparisons can be a problem, even when there is no ''fishing expedition'' or
  ''p-hacking'' and the research hypothesis was posited ahead of time}.
\newblock \bibinfo{journal}{\emph{Department of Statistics, Columbia
  University}} (\bibinfo{year}{2013}).
\newblock


\bibitem[\protect\citeauthoryear{Grinberg, Joseph, Friedland, Swire-Thompson,
  and Lazer}{Grinberg et~al\mbox{.}}{2019}]%
        {Grinberg2019}
\bibfield{author}{\bibinfo{person}{Nir Grinberg}, \bibinfo{person}{Kenneth
  Joseph}, \bibinfo{person}{Lisa Friedland}, \bibinfo{person}{Briony
  Swire-Thompson}, {and} \bibinfo{person}{David Lazer}.}
  \bibinfo{year}{2019}\natexlab{}.
\newblock \showarticletitle{{Fake news on Twitter during the 2016 U.S.
  presidential election}}.
\newblock \bibinfo{journal}{\emph{Science}} \bibinfo{volume}{363},
  \bibinfo{number}{6425} (\bibinfo{year}{2019}), \bibinfo{pages}{374--378}.
\newblock


\bibitem[\protect\citeauthoryear{Guess, Nagler, and Tucker}{Guess
  et~al\mbox{.}}{2019}]%
        {guess2019less}
\bibfield{author}{\bibinfo{person}{Andrew Guess}, \bibinfo{person}{Jonathan
  Nagler}, {and} \bibinfo{person}{Joshua Tucker}.}
  \bibinfo{year}{2019}\natexlab{}.
\newblock \showarticletitle{Less than you think: Prevalence and predictors of
  fake news dissemination on Facebook}.
\newblock \bibinfo{journal}{\emph{Science Advances}} \bibinfo{volume}{5},
  \bibinfo{number}{1} (\bibinfo{year}{2019}), \bibinfo{pages}{eaau4586}.
\newblock


\bibitem[\protect\citeauthoryear{Hajnal, Lajevardi, and Nielson}{Hajnal
  et~al\mbox{.}}{2017}]%
        {hajnal2017voter}
\bibfield{author}{\bibinfo{person}{Zoltan Hajnal}, \bibinfo{person}{Nazita
  Lajevardi}, {and} \bibinfo{person}{Lindsay Nielson}.}
  \bibinfo{year}{2017}\natexlab{}.
\newblock \showarticletitle{Voter identification laws and the suppression of
  minority votes}.
\newblock \bibinfo{journal}{\emph{The Journal of Politics}}
  \bibinfo{volume}{79}, \bibinfo{number}{2} (\bibinfo{year}{2017}).
\newblock


\bibitem[\protect\citeauthoryear{Highton}{Highton}{2004}]%
        {highton2004voter}
\bibfield{author}{\bibinfo{person}{Benjamin Highton}.}
  \bibinfo{year}{2004}\natexlab{}.
\newblock \showarticletitle{Voter registration and turnout in the United
  States}.
\newblock \bibinfo{journal}{\emph{Perspectives on Politics}}
  \bibinfo{volume}{2}, \bibinfo{number}{3} (\bibinfo{year}{2004}),
  \bibinfo{pages}{507--515}.
\newblock


\bibitem[\protect\citeauthoryear{Highton and Wolfinger}{Highton and
  Wolfinger}{1998}]%
        {highton1998estimating}
\bibfield{author}{\bibinfo{person}{Benjamin Highton} {and}
  \bibinfo{person}{Raymond~E Wolfinger}.} \bibinfo{year}{1998}\natexlab{}.
\newblock \showarticletitle{Estimating the effects of the National Voter
  Registration Act of 1993}.
\newblock \bibinfo{journal}{\emph{Political Behavior}} \bibinfo{volume}{20},
  \bibinfo{number}{2} (\bibinfo{year}{1998}), \bibinfo{pages}{79--104}.
\newblock


\bibitem[\protect\citeauthoryear{Howard, Bolsover, Kollanyi, Bradshaw, and
  Neudert}{Howard et~al\mbox{.}}{2017}]%
        {howard2017junk}
\bibfield{author}{\bibinfo{person}{Philip~N Howard}, \bibinfo{person}{Gillian
  Bolsover}, \bibinfo{person}{Bence Kollanyi}, \bibinfo{person}{Samantha
  Bradshaw}, {and} \bibinfo{person}{Lisa-Maria Neudert}.}
  \bibinfo{year}{2017}\natexlab{}.
\newblock \showarticletitle{Junk news and bots during the US election: What
  were Michigan voters sharing over Twitter}.
\newblock \bibinfo{journal}{\emph{CompProp, OII, Data Memo}}
  (\bibinfo{year}{2017}).
\newblock


\bibitem[\protect\citeauthoryear{Kim}{Kim}{2003}]%
        {kim2003help}
\bibfield{author}{\bibinfo{person}{Brian Kim}.}
  \bibinfo{year}{2003}\natexlab{}.
\newblock \bibinfo{title}{Help America Vote Act}.
\newblock
\newblock


\bibitem[\protect\citeauthoryear{Kudugunta and Ferrara}{Kudugunta and
  Ferrara}{2018}]%
        {kudugunta2018deep}
\bibfield{author}{\bibinfo{person}{Sneha Kudugunta} {and}
  \bibinfo{person}{Emilio Ferrara}.} \bibinfo{year}{2018}\natexlab{}.
\newblock \showarticletitle{Deep Neural Networks for Bot Detection}.
\newblock \bibinfo{journal}{\emph{Information Sciences}} \bibinfo{volume}{467},
  \bibinfo{number}{October} (\bibinfo{year}{2018}), \bibinfo{pages}{312--322}.
\newblock


\bibitem[\protect\citeauthoryear{Lazer, Kennedy, King, and Vespignani}{Lazer
  et~al\mbox{.}}{2014}]%
        {lazer2014parable}
\bibfield{author}{\bibinfo{person}{David Lazer}, \bibinfo{person}{Ryan
  Kennedy}, \bibinfo{person}{Gary King}, {and} \bibinfo{person}{Alessandro
  Vespignani}.} \bibinfo{year}{2014}\natexlab{}.
\newblock \showarticletitle{The parable of Google Flu: traps in big data
  analysis}.
\newblock \bibinfo{journal}{\emph{Science}} \bibinfo{volume}{343},
  \bibinfo{number}{6176} (\bibinfo{year}{2014}).
\newblock


\bibitem[\protect\citeauthoryear{Malik, Lamba, Nakos, and Pfeffer}{Malik
  et~al\mbox{.}}{2015}]%
        {malik2015population}
\bibfield{author}{\bibinfo{person}{Momin~M Malik}, \bibinfo{person}{Hemank
  Lamba}, \bibinfo{person}{Constantine Nakos}, {and} \bibinfo{person}{Jurgen
  Pfeffer}.} \bibinfo{year}{2015}\natexlab{}.
\newblock \showarticletitle{Population bias in geotagged tweets}.
\newblock \bibinfo{journal}{\emph{People}} \bibinfo{volume}{1},
  \bibinfo{number}{3,759.710} (\bibinfo{year}{2015}), \bibinfo{pages}{3--759}.
\newblock


\bibitem[\protect\citeauthoryear{Metaxas, Mustafaraj, and Gayo-Avello}{Metaxas
  et~al\mbox{.}}{2011}]%
        {metaxas2011not}
\bibfield{author}{\bibinfo{person}{Panagiotis~T Metaxas}, \bibinfo{person}{Eni
  Mustafaraj}, {and} \bibinfo{person}{Dani Gayo-Avello}.}
  \bibinfo{year}{2011}\natexlab{}.
\newblock \showarticletitle{How (not) to predict elections}. In
  \bibinfo{booktitle}{\emph{3rd International Conference on Social Computing}}.
  \bibinfo{pages}{165--171}.
\newblock


\bibitem[\protect\citeauthoryear{Minnite}{Minnite}{2017}]%
        {minnite2017myth}
\bibfield{author}{\bibinfo{person}{Lorraine~C Minnite}.}
  \bibinfo{year}{2017}\natexlab{}.
\newblock \bibinfo{booktitle}{\emph{The myth of voter fraud}}.
\newblock \bibinfo{publisher}{Cornell University Press}.
\newblock


\bibitem[\protect\citeauthoryear{Mislove, Lehmann, Ahn, Onnela, and
  Rosenquist}{Mislove et~al\mbox{.}}{2011}]%
        {mislove2011understanding}
\bibfield{author}{\bibinfo{person}{Alan Mislove}, \bibinfo{person}{Sune
  Lehmann}, \bibinfo{person}{Yong-Yeol Ahn}, \bibinfo{person}{Jukka-Pekka
  Onnela}, {and} \bibinfo{person}{J~Niels Rosenquist}.}
  \bibinfo{year}{2011}\natexlab{}.
\newblock \showarticletitle{Understanding the Demographics of Twitter Users.}
\newblock \bibinfo{journal}{\emph{ICWSM}} (\bibinfo{year}{2011}).
\newblock


\bibitem[\protect\citeauthoryear{M{\o}nsted, Sapie{\.z}y{\'n}ski, Ferrara, and
  Lehmann}{M{\o}nsted et~al\mbox{.}}{2017}]%
        {monsted2017evidence}
\bibfield{author}{\bibinfo{person}{Bjarke M{\o}nsted}, \bibinfo{person}{Piotr
  Sapie{\.z}y{\'n}ski}, \bibinfo{person}{Emilio Ferrara}, {and}
  \bibinfo{person}{Sune Lehmann}.} \bibinfo{year}{2017}\natexlab{}.
\newblock \showarticletitle{Evidence of Complex Contagion of Information in
  Social Media: An Experiment Using Twitter Bots}.
\newblock \bibinfo{journal}{\emph{Plos One}} \bibinfo{volume}{12},
  \bibinfo{number}{9} (\bibinfo{year}{2017}), \bibinfo{pages}{e0184148}.
\newblock


\bibitem[\protect\citeauthoryear{Morstatter, Pfeffer, Liu, and
  Carley}{Morstatter et~al\mbox{.}}{2013}]%
        {morstatter2013sample}
\bibfield{author}{\bibinfo{person}{Fred Morstatter},
  \bibinfo{person}{J{\"u}rgen Pfeffer}, \bibinfo{person}{Huan Liu}, {and}
  \bibinfo{person}{Kathleen~M Carley}.} \bibinfo{year}{2013}\natexlab{}.
\newblock \showarticletitle{Is the Sample Good Enough? Comparing Data from
  Twitter's Streaming API with Twitter's Firehose}. In
  \bibinfo{booktitle}{\emph{ICWSM}}.
\newblock


\bibitem[\protect\citeauthoryear{Mosteller and Doob}{Mosteller and
  Doob}{1949}]%
        {mosteller1949pre}
\bibfield{author}{\bibinfo{person}{Frederick Mosteller} {and}
  \bibinfo{person}{Leonard~William Doob}.} \bibinfo{year}{1949}\natexlab{}.
\newblock \bibinfo{booktitle}{\emph{The pre-election polls of 1948}}.
\newblock \bibinfo{publisher}{Social Science Research Council}.
\newblock


\bibitem[\protect\citeauthoryear{Pennycook and Rand}{Pennycook and
  Rand}{2018}]%
        {PENNYCOOK2018}
\bibfield{author}{\bibinfo{person}{Gordon Pennycook} {and}
  \bibinfo{person}{David~G. Rand}.} \bibinfo{year}{2018}\natexlab{}.
\newblock \showarticletitle{Lazy, not biased: Susceptibility to partisan fake
  news is better explained by lack of reasoning than by motivated reasoning}.
\newblock \bibinfo{journal}{\emph{Cognition}} (\bibinfo{year}{2018}).
\newblock


\bibitem[\protect\citeauthoryear{Pennycook and Rand}{Pennycook and
  Rand}{2019}]%
        {pennycook2019cognitive}
\bibfield{author}{\bibinfo{person}{Gordon Pennycook} {and}
  \bibinfo{person}{David~G Rand}.} \bibinfo{year}{2019}\natexlab{}.
\newblock \showarticletitle{Cognitive reflection and the 2016 US presidential
  election}.
\newblock \bibinfo{journal}{\emph{Personality and Social Psychology Bulletin}}
  \bibinfo{volume}{45}, \bibinfo{number}{2} (\bibinfo{year}{2019}).
\newblock


\bibitem[\protect\citeauthoryear{Persily}{Persily}{2017}]%
        {persily20172016}
\bibfield{author}{\bibinfo{person}{Nathaniel Persily}.}
  \bibinfo{year}{2017}\natexlab{}.
\newblock \showarticletitle{The 2016 US Election: Can democracy survive the
  internet?}
\newblock \bibinfo{journal}{\emph{Journal of democracy}} \bibinfo{volume}{28},
  \bibinfo{number}{2} (\bibinfo{year}{2017}), \bibinfo{pages}{63--76}.
\newblock


\bibitem[\protect\citeauthoryear{Pitoura, Tsaparas, Flouris, Fundulaki,
  Papadakos, Abiteboul, and Weikum}{Pitoura et~al\mbox{.}}{2018}]%
        {pitoura2018measuring}
\bibfield{author}{\bibinfo{person}{Evaggelia Pitoura},
  \bibinfo{person}{Panayiotis Tsaparas}, \bibinfo{person}{Giorgos Flouris},
  \bibinfo{person}{Irini Fundulaki}, \bibinfo{person}{Panagiotis Papadakos},
  \bibinfo{person}{Serge Abiteboul}, {and} \bibinfo{person}{Gerhard Weikum}.}
  \bibinfo{year}{2018}\natexlab{}.
\newblock \showarticletitle{On Measuring Bias in Online Information}.
\newblock \bibinfo{journal}{\emph{ACM SIGMOD Record}} \bibinfo{volume}{46},
  \bibinfo{number}{4} (\bibinfo{year}{2018}), \bibinfo{pages}{16--21}.
\newblock


\bibitem[\protect\citeauthoryear{Ruths and Pfeffer}{Ruths and Pfeffer}{2014}]%
        {ruths2014social}
\bibfield{author}{\bibinfo{person}{Derek Ruths} {and}
  \bibinfo{person}{J{\"u}rgen Pfeffer}.} \bibinfo{year}{2014}\natexlab{}.
\newblock \showarticletitle{Social media for large studies of behavior}.
\newblock \bibinfo{journal}{\emph{Science}} \bibinfo{volume}{346},
  \bibinfo{number}{6213} (\bibinfo{year}{2014}), \bibinfo{pages}{1063--1064}.
\newblock


\bibitem[\protect\citeauthoryear{Scheufele and Krause}{Scheufele and
  Krause}{2019}]%
        {scheufele2019science}
\bibfield{author}{\bibinfo{person}{Dietram~A Scheufele} {and}
  \bibinfo{person}{Nicole~M Krause}.} \bibinfo{year}{2019}\natexlab{}.
\newblock \showarticletitle{Science audiences, misinformation, and fake news}.
\newblock \bibinfo{journal}{\emph{PNAS}} (\bibinfo{year}{2019}),
  \bibinfo{pages}{201805871}.
\newblock


\bibitem[\protect\citeauthoryear{Shao, Ciampaglia, Varol, Yang, Flammini, and
  Menczer}{Shao et~al\mbox{.}}{2018}]%
        {shao2018spread}
\bibfield{author}{\bibinfo{person}{Chengcheng Shao},
  \bibinfo{person}{Giovanni~Luca Ciampaglia}, \bibinfo{person}{Onur Varol},
  \bibinfo{person}{Kai-Cheng Yang}, \bibinfo{person}{Alessandro Flammini},
  {and} \bibinfo{person}{Filippo Menczer}.} \bibinfo{year}{2018}\natexlab{}.
\newblock \showarticletitle{The spread of low-credibility content by social
  bots}.
\newblock \bibinfo{journal}{\emph{Nature communications}} \bibinfo{volume}{9},
  \bibinfo{number}{1} (\bibinfo{year}{2018}), \bibinfo{pages}{4787}.
\newblock


\bibitem[\protect\citeauthoryear{Shu, Sliva, Wang, Tang, and Liu}{Shu
  et~al\mbox{.}}{2017}]%
        {shu2017fake}
\bibfield{author}{\bibinfo{person}{Kai Shu}, \bibinfo{person}{Amy Sliva},
  \bibinfo{person}{Suhang Wang}, \bibinfo{person}{Jiliang Tang}, {and}
  \bibinfo{person}{Huan Liu}.} \bibinfo{year}{2017}\natexlab{}.
\newblock \showarticletitle{Fake news detection on social media: A data mining
  perspective}.
\newblock \bibinfo{journal}{\emph{ACM SIGKDD Explorations Newsletter}}
  \bibinfo{volume}{19}, \bibinfo{number}{1} (\bibinfo{year}{2017}),
  \bibinfo{pages}{22--36}.
\newblock


\bibitem[\protect\citeauthoryear{Stella, Ferrara, and De~Domenico}{Stella
  et~al\mbox{.}}{2018}]%
        {stella2018bots}
\bibfield{author}{\bibinfo{person}{Massimo Stella}, \bibinfo{person}{Emilio
  Ferrara}, {and} \bibinfo{person}{Manlio De~Domenico}.}
  \bibinfo{year}{2018}\natexlab{}.
\newblock \showarticletitle{Bots increase exposure to negative and inflammatory
  content in online social systems}.
\newblock \bibinfo{journal}{\emph{Proceedings of the National Academy of
  Sciences}} \bibinfo{volume}{115}, \bibinfo{number}{49}
  (\bibinfo{year}{2018}), \bibinfo{pages}{12435--12440}.
\newblock


\bibitem[\protect\citeauthoryear{Varol, Ferrara, Davis, Menczer, and
  Flammini}{Varol et~al\mbox{.}}{2017a}]%
        {varol2017online}
\bibfield{author}{\bibinfo{person}{Onur Varol}, \bibinfo{person}{Emilio
  Ferrara}, \bibinfo{person}{Clayton~A Davis}, \bibinfo{person}{Filippo
  Menczer}, {and} \bibinfo{person}{Alessandro Flammini}.}
  \bibinfo{year}{2017}\natexlab{a}.
\newblock \showarticletitle{Online human-bot interactions: Detection,
  estimation, and characterization}. In \bibinfo{booktitle}{\emph{Int. AAAI
  Conference on Web and Social Media}}. \bibinfo{pages}{280--289}.
\newblock


\bibitem[\protect\citeauthoryear{Varol, Ferrara, Menczer, and Flammini}{Varol
  et~al\mbox{.}}{2017b}]%
        {varol2017early}
\bibfield{author}{\bibinfo{person}{Onur Varol}, \bibinfo{person}{Emilio
  Ferrara}, \bibinfo{person}{Filippo Menczer}, {and}
  \bibinfo{person}{Alessandro Flammini}.} \bibinfo{year}{2017}\natexlab{b}.
\newblock \showarticletitle{Early Detection of Promoted Campaigns on Social
  Media}.
\newblock \bibinfo{journal}{\emph{EPJ Data Science}} \bibinfo{volume}{6},
  \bibinfo{number}{13} (\bibinfo{year}{2017}).
\newblock


\bibitem[\protect\citeauthoryear{Vosoughi, Roy, and Aral}{Vosoughi
  et~al\mbox{.}}{2018}]%
        {vosoughi2018spread}
\bibfield{author}{\bibinfo{person}{Soroush Vosoughi}, \bibinfo{person}{Deb
  Roy}, {and} \bibinfo{person}{Sinan Aral}.} \bibinfo{year}{2018}\natexlab{}.
\newblock \showarticletitle{The spread of true and false news online}.
\newblock \bibinfo{journal}{\emph{Science}} \bibinfo{volume}{359},
  \bibinfo{number}{6380} (\bibinfo{year}{2018}), \bibinfo{pages}{1146--1151}.
\newblock


\bibitem[\protect\citeauthoryear{Wang}{Wang}{2012}]%
        {wang2012politics}
\bibfield{author}{\bibinfo{person}{Tova Wang}.}
  \bibinfo{year}{2012}\natexlab{}.
\newblock \bibinfo{booktitle}{\emph{The politics of voter suppression:
  Defending and expanding Americans' right to vote}}.
\newblock \bibinfo{publisher}{Cornell University Press}.
\newblock


\bibitem[\protect\citeauthoryear{Williams, Burnap, and Sloan}{Williams
  et~al\mbox{.}}{2017}]%
        {williams2017crime}
\bibfield{author}{\bibinfo{person}{Matthew~L Williams}, \bibinfo{person}{Pete
  Burnap}, {and} \bibinfo{person}{Luke Sloan}.}
  \bibinfo{year}{2017}\natexlab{}.
\newblock \showarticletitle{Crime sensing with big data: The affordances and
  limitations of using open-source communications to estimate crime patterns}.
\newblock \bibinfo{journal}{\emph{The British Journal of Criminology}}
  \bibinfo{volume}{57}, \bibinfo{number}{2} (\bibinfo{year}{2017}).
\newblock


\bibitem[\protect\citeauthoryear{Woolley and Guilbeault}{Woolley and
  Guilbeault}{2017}]%
        {woolley2017computational}
\bibfield{author}{\bibinfo{person}{Samuel~C Woolley} {and}
  \bibinfo{person}{Douglas~R Guilbeault}.} \bibinfo{year}{2017}\natexlab{}.
\newblock \showarticletitle{Computational propaganda in the United States of
  America: Manufacturing consensus online}.
\newblock \bibinfo{journal}{\emph{Computational Propaganda Research Project}}
  (\bibinfo{year}{2017}), \bibinfo{pages}{22}.
\newblock


\bibitem[\protect\citeauthoryear{Yang, Varol, Davis, Ferrara, Flammini, and
  Menczer}{Yang et~al\mbox{.}}{2019}]%
        {yang2019arming}
\bibfield{author}{\bibinfo{person}{Kai-Cheng Yang}, \bibinfo{person}{Onur
  Varol}, \bibinfo{person}{Clayton~A Davis}, \bibinfo{person}{Emilio Ferrara},
  \bibinfo{person}{Alessandro Flammini}, {and} \bibinfo{person}{Filippo
  Menczer}.} \bibinfo{year}{2019}\natexlab{}.
\newblock \showarticletitle{Arming the public with AI to counter social bots}.
\newblock \bibinfo{journal}{\emph{Human Behavior and Emerging Technologies}}
  \bibinfo{volume}{1}, \bibinfo{number}{1} (\bibinfo{year}{2019}).
\newblock


\end{thebibliography}
